\pdfoutput=1
\documentclass[a4paper,fleqn,usenatbib]{mnras}

\usepackage{newtxtext,newtxmath}

\usepackage[T1]{fontenc}
\usepackage{ae,aecompl}

\usepackage{graphicx}	
\usepackage{amsmath}	
\usepackage{amssymb}	
\usepackage{subfigure}
\usepackage{float}
\usepackage{changepage}
\usepackage{url}
\usepackage{rotating}
\usepackage{threeparttable}
\usepackage{caption}
\usepackage{indentfirst}

\title[Galaxy shape evolution: prolate to oblate]{The evolution of galaxy shapes in CANDELS: from prolate to oblate}

\author[H. Zhang et al.]{
Haowen Zhang,$^{1,2}$\thanks{E-mail: zhw0595@pku.edu.cn}
Joel R. Primack,$^{3,4}$
S. M. Faber,$^{5}$
David C. Koo,$^{5}$
Avishai Dekel,$^{6,4}$
\newauthor{
Zhu Chen,$^{7}$
Daniel Ceverino,$^{8}$
Yu-Yen Chang,$^{9}$
Jerome J. Fang,$^{10}$
Yicheng Guo,$^{11}$
}
\newauthor{
Lin Lin$^{12}$
and Arjen van der Wel$^{13,14}$
}
\\
$^{1}$The School of Physics, Peking University, Peking University, Beijing 100871, China\\
$^{2}$Kavli Institute of Astronomy and Astrophysics, Peking University, Peking University, Beijing 100871, China\\
$^{3}$Physics Department, University of California, Santa Cruz, CA 95064 USA\\
$^{4}$SCIPP, University of California, Santa Cruz, CA 95064 USA\\
$^{5}$UCO/Lick Observatory, Dept. of Astronomy and Astrophysics, University of California, Santa Cruz, CA 95064 USA\\
$^{6}$Racah Institute of Physics, The Hebrew University, Jerusalem 91904, Israel\\
$^{7}$Mathematics and Science College of Shanghai Normal University, Shanghai, China\\
$^{8}$Zentrum f\"ur Astronomie der Universit\"at Heidelberg, Institut f\"ur Theoretische Astrophysik, Germany\\
$^{9}$Academia Sinica Institute of Astronomy and Astrophysics, PO Box 23-141, Taipei 10617, Taiwan\\
$^{10}$Orange Coast College, Costa Mesa, CA 92626, USA\\
$^{11}$Dept. of Physics and Astronomy, University of Missouri, Columbia, MO 65211-7010\\
$^{12}$Shanghai Astronomical Observatory, Shanghai, China\\
$^{13}$Sterrenkundig Observatorium, Universiteit Gent, Krijgslaan 281 S9, B-9000 Gent, Belgium\\
$^{14}$Max-Planck Institut f\"ur Astronomie, K\"onigstuhl 17, D- 69117, Heidelberg, Germany
}

\date{Accepted XXX. Received YYY; in original form ZZZ}

\pubyear{2018}

\begin{document}
\label{firstpage}
\pagerange{\pageref{firstpage}--\pageref{lastpage}}
\maketitle
\bibpunct{(}{)}{;}{a}{}{,}

\begin{abstract}
We model the projected $b/a - \log a$ distributions of CANDELS main sequence star-forming galaxies, where $a$ ($b$) is the semi-major (semi-minor) axis of the galaxy images. We find that smaller-$a$ galaxies are rounder at all stellar masses $M_{*}$ and redshifts, so we include $a$ when analyzing $b/a$ distributions. Approximating intrinsic shapes of the galaxies as triaxial ellipsoids and assuming a multivariate normal distribution of galaxy size and two shape parameters, we construct their intrinsic shape and size distributions to obtain the fractions of prolate, oblate, and spheroidal galaxies in each redshift and mass bin. We find that galaxies tend to be prolate at low $M_{*}$  and high redshifts, and oblate at high $M_{*}$  and low redshifts, qualitatively consistent with van der Wel et al. (2014), implying that galaxies tend to evolve from prolate to oblate. These results are consistent with the predictions from simulations (Ceverino et al. 2015, Tomassetti et al. 2016) that the transition from prolate to oblate is caused by a compaction event at a characteristic mass range, making the galaxy center baryon dominated. We give probabilities of a galaxy's being prolate, oblate, or spheroidal as a function of its $M_{*}$, redshift, and projected $b/a$ and $a$, which can facilitate target selections of galaxies with specific shapes at high redshifts. We also give predicted optical depths of galaxies, which are qualitatively consistent with the expected correlation that $A_{\rm V}$ should be higher for edge-on disk galaxies in each $\log a$ slice at low redshift and high mass bins. 
\end{abstract}




\section{Introduction}

The shape of a galaxy's stellar component is closely related to the formation and evolution of the galaxy. Although the effect of viewing orientations plays some role, the evolution of the projected shapes of galaxies reflect that of their intrinsic shapes. \citet{Elmegreen:2005hs} found that the projected ellipticity distribution of the spiral galaxies in the Hubble Ultra Deep Field peaks (HUDF) at $\sim0.55$, indicating thicker disks by a factor of $\sim2$ than the local disk galaxies. Similar findings have been made on the Lyman Break Galaxies (LBGs) in the Great Observatories Origins Deep Survey (GOODS) by \citet{Ravindranath:2006ie}, which finds an evolution of the peak of the projected ellipicity from $\sim0.7$ at $z=4$ to $\sim0.5$ at $z=3$. Later works went further by modeling the intrinsic shape distributions of galaxies using the projected $b/a$ axis ratio distributions. By assuming that the shapes of ellipticals and spirals can be well approximated by triaxial ellipsoids, \citet{Padilla:2008gw} modeled the intrinsic shape distribution of a subset of SDSS galaxies, and found that generally brighter (and more massive) galaxies tend to be rounder. They also found that the intrinsic shapes of the galaxies are correlated with their sizes in the sense that the median $b/a$ decreases as one looks at the galaxies with larger $a$. 

Based on similar modeling methodology, \citet{Law:2012hf} found that the intrinsic shapes of the star-forming galaxies in the HST/WFC3 survey with $1.5 < z < 3.6$ are more consistent with a triaxial population, with intrinsic $b/a\sim0.7$ and $c/a\sim0.3$, rather than thick oblate disks despite the galaxies mostly having exponential surface brightness profiles (S\'{e}rsic index $n \sim 1$). \Citet{vanderWel:2014ka} determined the intrinsic shape distributions of star-forming galaxies with $0 < z < 2.5$ from SDSS and CANDELS. The basic assumption was that the galaxies are triaxial ellipsoids with axes $a \geq b \geq c$ with a Gaussian distribution of the ellipticity ($E = 1 - c/a$) and triaxiality ($T = \left(a^2 - b^2\right) / \left(a^2 - c^2\right)$). By finding the best fitting parameters that describe such Gaussian distributions, they found that the fraction of prolate galaxies decreases with increasing time and mass. At high redshift, they found that low mass galaxies are a mixture of roughly equal numbers of prolate and oblate galaxies, while the fraction of prolate galaxies remains negligible for the most massive populations throughout the whole redshift range. Qualitatively the picture indicated by their results is that the overall oblateness (i.e. the fractions of oblate objects) of galaxies increases with time, and that this process proceeds earlier in higher-mass galaxies. Recently, \citet{Jiang2018} analyzed the radial profiles of the isophotal ellipticity $\epsilon=1-b/a$ and disky/boxy parameter $A_4$ of $\sim$4600 star-forming galaxies (SFGs) with  in CANDELS with $9.0 < \log \left( M_{*}/M_\odot\right) < 11.0$ and $0.5 < z < 1.8$. By dividing the whole sample into a series of redshift-mass bins, they found that the more massive galaxies in lower redshifts have more disky isophotes at intermediate radii, which is consistent with the picture indicated by the findings of \citet{vanderWel:2014ka}. By further dividing the sample into large and small SFGs using the deviation from the size-mass relation in each redshift bin, they also found that larger SFGs typically have isophotes with larger $\epsilon$ (i.e. rounder) when compared with small SFGs.

The evolution of galaxy shapes has also been investigated from a theoretical perspective. A shape evolution from prolate to oblate in three dimensions has been revealed and studied in the VELA zoom-in cosmological simulations \citep{Ceverino:2015db,Tomassetti:2015ed}.  In the simulations, the galaxies tend to experience an early prolate phase while their interiors (radii lower than the galaxy half-mass radius) are dominated by dark matter. They thus follow the prolateness of the inner dark-matter halo, generated by mergers along a cosmic-web filament and being supported by anisotropic velocity dispersion \citep{Allgood:2006bb}. After a major wet compaction event in which inflowing gas makes the galaxy compact and baryon dominated, the stellar orbits supporting the prolateness are deflected and the system evolves into an oblate shape, induced by the angular-momentum of the newly accreted mass. The major compaction events tend to occcur in a characteristic mass range, $M_*\sim10^{9.5-10} M_{\mathrm{\odot}}$ \citep{Zolotov:2015hk,Tomassetti:2015ed,Tacchella:2016a,Tacchella:2016b}. Therefore, more massive galaxies at a given redshift are predicted to make the transition from prolate to oblate at a higher redshift. One can test these predictions using the projected shapes of a large observed sample of galaxies.

There exists a potential problem with several of the previous observational analyses, which were based on the projected $b/a$ distribution only: as \citet{Padilla:2008gw} pointed out, the projected and intrinsic shapes of SDSS galaxies are correlated with their $R_e$ (hereafter $a$). Similarly, \citet{FangThesis} and \citet{Fang2017} found that the projected shapes of the CANDELS galaxies are correlated with their residuals from the $\log a-\log M_{*}$ relation, which serves as another indicator of galaxy sizes. Given this fact, one should in principle carry out such $b/a$ modelings on galaxy subsamples in different $a$ bins as \citet{Padilla:2008gw} did to avoid potential bias in the fractions of the galaxies with different shapes. Alternatively one can also directly model the two-dimensional projected $b/a-\log a$ distribution and overcome this potential bias by allowing correlation between intrinsic shape parameters and sizes. In this work we try to make such two-dimensional modelings to better determine the fractions of star-forming galaxies with different shapes at a given mass and redshift, and to further test the picture that the oblateness (prolateness) of star-forming galaxies increases (decreases) with increasing time and mass, which is expected, if the prediction made by \citet{Ceverino:2015db} and \citet{Tomassetti:2015ed} is true, that there is a characteristic mass range where the galaxies go through a shape transition.

This paper is organized as follows. In Section 2, we describe our criteria for data selection, and give some visual impressions of the distributions of the data in the projected $b/a-\log a$ plane. Section 3 describes our methodology to model such a two-dimensional distribution. The modeling results are shown in Section 4. In Section 5 we give some further discussions based on our modeling results. Section 6 lists several caveats regarding the analysis in this work. In Section 7 we summarize our results.  Appendices discuss tests of potential selection effects in the CANDELS pipeline and provide further discussions of our modeling of galaxy shapes.

Throughout this paper we use AB magnitudes and adopt the cosmological parameters $\left(\Omega_{\mathrm{M}}, \Omega_{\mathrm{\Lambda}}, h\right)=(0.3, 0.7, 0.7)$.

\section{Data} 
\label{sec:data}

This work makes use of the multi-wavelength and ancillary datasets produced by the Cosmic Assembly Near-Infrared Deep Extragalactic Legacy Survey \citep[CANDELS: ][]{2011ApJS..197...35G,2011ApJS..197...36K}. The data reduction and cataloging for each of the fields is presented in \citet[COSMOS]{Nayyeri:2017jd}, \citet[EGS]{Stefanon:2017dm}, Barro et al. (in prep., GOODS-N), \citet[GOODS-S]{Guo:2013ig,Santini:2015hh} and \citet[UDS]{Galametz:2013dd,Santini:2015hh}.
 The \textsc{GALFIT} measurements of the $b/a$ and $a$ of CANDELS galaxies used in this work are obtained by \citet{2012ApJS..203...24V}\footnote{All of the data used in this work can be accessed from the \textit{Rainbow} database \citep{Barro:2011is}, whose URL is \url{http://arcoiris.ucsc.edu//Rainbow\_navigator\_public/}. All the \textsc{GALFIT} output files can be found at \url{http://www.mpia.de/homes/vdwel/3dhstcandels.html}.}.

\subsection{Data selection} 
\label{sub:data_selection}{}

For the sake of a direct comparison between this work and \citet{vanderWel:2014ka}, it would have been best to model the same data that they used. But \citet{vanderWel:2014ka} used the UVJ diagram \citep{Wuyts:2007hs,Williams:2009hn} method to pick out star-forming galaxies, which tends to include green valley galaxies. Also, their sample is highly incomplete in the $2.0 < z < 2.5$ and $9.0 < \log \left( M_{*}/M_\odot\right) < 9.5$ bin. Thus in addition to the galaxies used in \citet{vanderWel:2014ka}, we also pick out just the star-forming main-sequence (SFMS) galaxies in CANDELS data, which is a complete sample in all the redshift and mass bins we study. Since this different sample selection doesn't change the main conclusions of the present work, we will only show the results based on the SFMS galaxies in CANDELS.

Our selection criteria for SFMS galaxies are based on the deviation from the star-forming main sequence defined by specific star formation rates ($\Delta \mathrm{SSFR}$) in each redshift-mass bin following the same method and formulae of \citet{Fang2017}. The SSFR-mass main sequence in each redshift bin is defined as \citep{Fang2017}:

\begin{equation}
\langle\log \mathrm{SSFR}\rangle = c_1(\log M_{*} - 10) + c_2\ ,
\end{equation}
where the parameters $c_1$ and $c_2$ are determined in the linear fitting. Based on $\Delta \mathrm{SSFR}$, our selection criteria are:

\begin{itemize}
\item[(1)] $H_{\mathrm{F160W}} < 25.5$;
\item[(2)] $z < 2.5$;
\item[(3)] $M_{*} > 10^9M_\odot$;
\item[(4)] $\Delta \mathrm{SSFR} > -0.45\ \mathrm{dex}$;
\item[(5)] have good \textsc{GALFIT} measurements of the structural parameters.
\end{itemize}

The data used in our modeling are the projected axis ratio $b/a$ and projected semi-major axis $a$ of the observed galaxies. These structural parameters are measured by \textsc{GALFIT} \citep{Peng:2010eh}, and the \textsc{GALFIT} setup can be found in \citet{2012ApJS..203...24V}. To have the galaxies' shapes measured at a rest-frame wavelength as close as possible to 4600\AA\ in order to avoid the effect that the shape of a galaxy changes with wavelength, which is seen in local galaxies \citep{Dalcanton:2002ga}, we use the structural parameters measured from H band (F160W) images for $2 < z < 2.5$ galaxies and the ones measured from J band (F125W) images for those with $z < 2$. 

The lowest redshift bins ($0 < z < 0.5$) and the most massive bins ($10.5 < \log \left(M_{*} / M_\odot\right) < 11.0$) in \citet{vanderWel:2014ka} are excluded from the quantitative modeling because our two-dimensional binning process requires a minimum sample size, and the numbers of the galaxies in such bins are too small to be modeled robustly. But we will still give some qualitative comments on the general trends with redshift shown in the most massive bins. Note that the numbers and the histograms in \textbf{Fig. 1} of \citet{vanderWel:2014ka} don't match each other: the numbers in the first row correspond to the histograms in the second row, and vice versa. We refer the readers to \citet{2012ApJS..203...24V} for more information on the determination of the measurement uncertainties of the structural parameters. 


\subsection{Visual impressions of the data} 
\label{sub:visual_impressions_on_the_data}

\begin{figure*}
\centering
\includegraphics[width=1.0\textwidth]{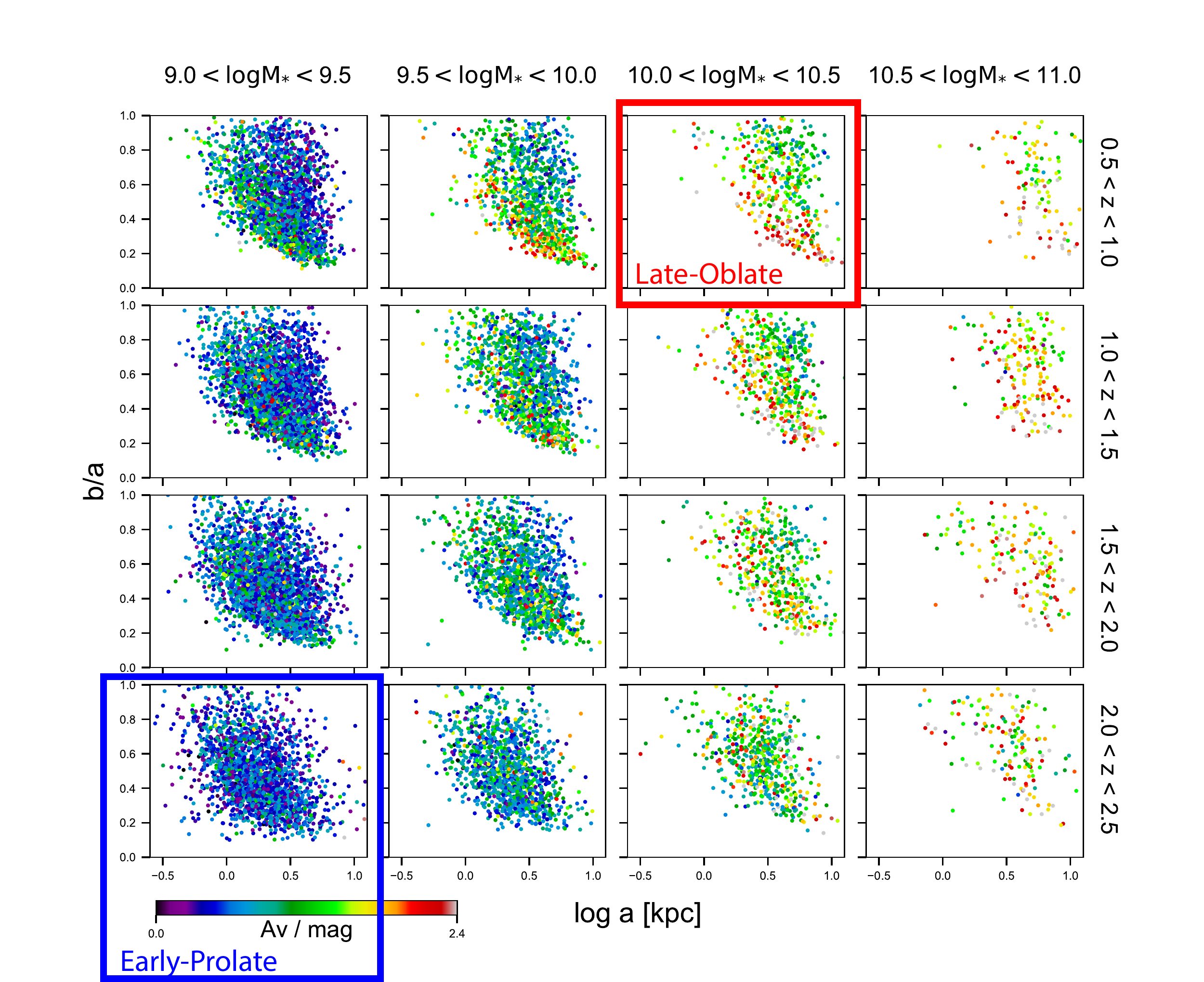}
\caption{The projected $b/a-\log a$ distributions of CANDELS galaxies in all the redshift-mass bins.  Here stellar mass $M_\ast$ is in units of $M_\odot$, and the points are color-coded by attenuation $A_V$ values from SED fitting. Note the strong trend of smaller galaxies being rounder and the increasing number of galaxies in the upper right corner of the panels with increasing time and mass.}
\label{fig1}
\end{figure*}

\textbf{Fig. 1} shows the projected $b/a-\log a$ scatter plot for galaxies in each bins of mass and redshift, color-coded by their $A_V$ values.  Note that galaxies evolve with redshift diagonally upwards and to the right in this diagram \citep[][fig. 5]{Fang2017}. There are some noteworthy features in these plots. Firstly a curved boundary at the lower left corner of each panel is clearly seen, which is qualitatively compatible with a population of galaxies with a constant or slowly evolving intrinsic shortest main axis. Secondly there's a lack of objects at the upper right corner of each panel at higher redshifts, increasingly filled with galaxies as the redshift decreases and the mass increases. We will give a more extended discussion of this trend with redshift and mass in Section 4.1. But we emphasize that these two features are not induced by the selection effects in the detection and measurement pipeline. To demonstrate this, we carried out a two-step experiment, each step of which mocks the process of detections and measurements in the pipeline, respectively. See the Appendix for a description of the assumptions, implementation, and results of such an experiment. Given the visual impressions of the $b/a-\log a$ distributions of the galaxies, one can be convinced that the assumption of \citet{vanderWel:2014ka} that the intrinsic shapes of galaxies are independent of their sizes is in fact not the case.  The implications of the $A_V$ distributions are discussed in Sections 4 and 5.

\begin{figure*}
\centering
\includegraphics[width=1.0\textwidth]{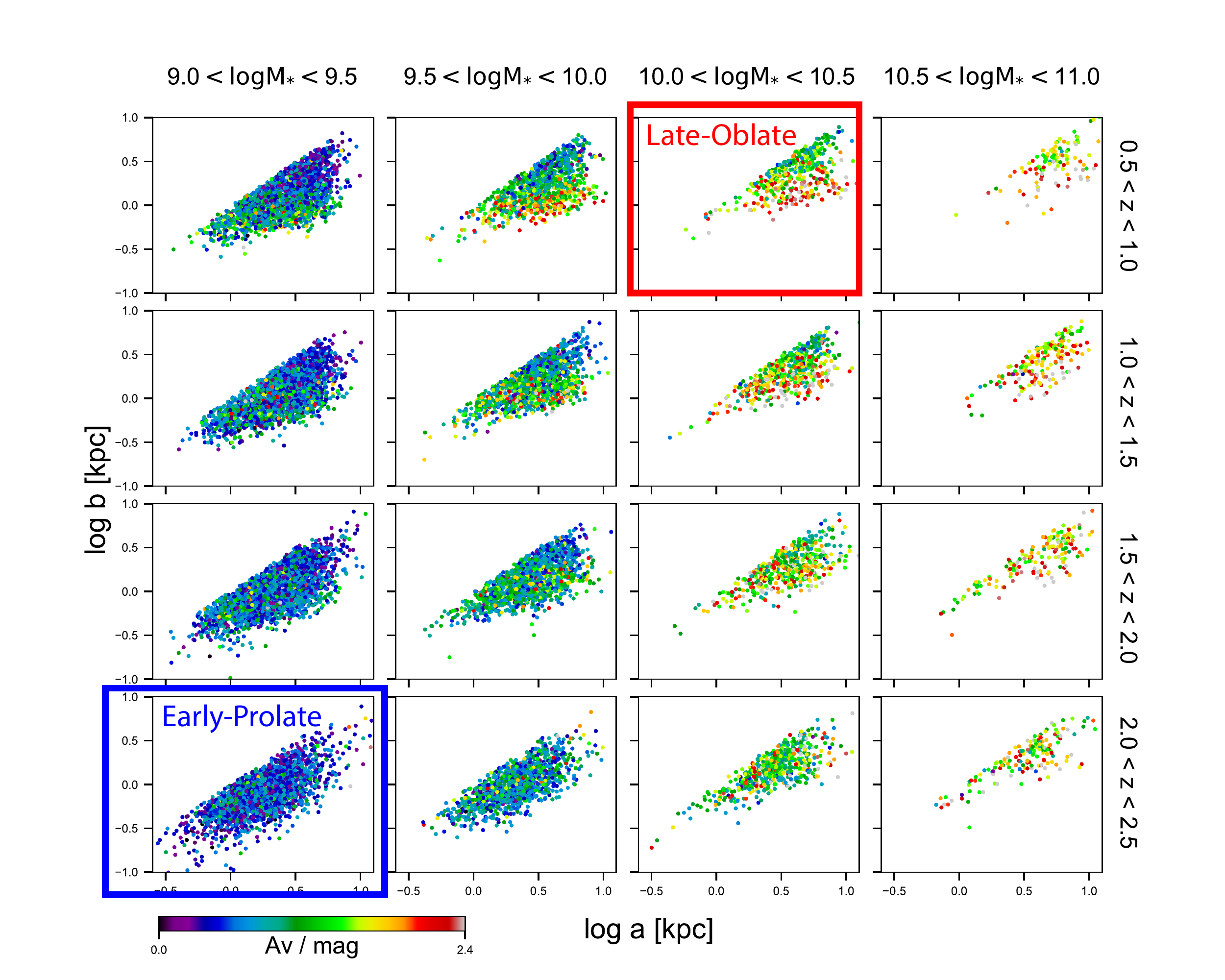}
\caption{$\log b - \log a$ distributions of the galaxies in all the redshift-mass bins, color-coded by $A_V$ values from SED fitting.}
\label{fig2}
\end{figure*}

\textbf{Fig. 2} shows the projected $\log b - \log a$ distributions of the galaxies with different masses and redshifts, again color-coded by $A_V$ values. We can see that in most panels the lower boundary of the distribution can be fitted pretty well by a straight line, which may imply relations between the intrinsic $c$ and $a$. This changing lower boundary with $\log a$ serves as further supporting evidence that the curved boundary giving the smallest $b/a$ is not an artificial result of the instrumental resolutions, because if the opposite were the case, the lowest observed $\log b$ would be constant over the whole range of $\log a$.

\begin{figure*}
\centering
\includegraphics[width=1.0\textwidth]{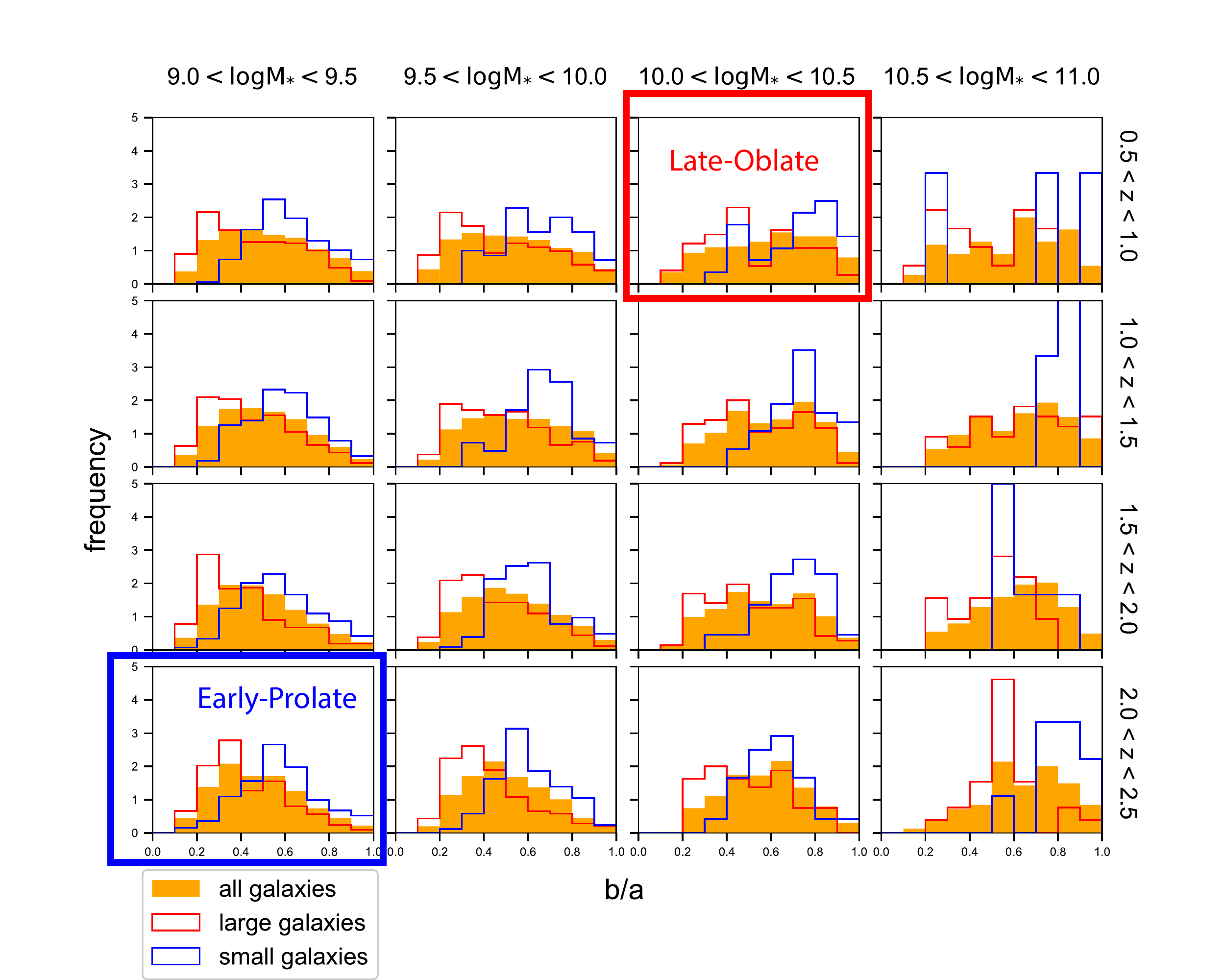}
\caption{The normalized $b/a$ histograms of galaxies of different sizes in all the redshift-mass bins. Red open histogram: large galaxies with $0.15 < \Delta \log a < 0.25$, with $a$ in kpc. Blue open histogram: small galaxies with $-0.35 < \Delta \log a < -0.25$.  Orange filled histogram: all galaxies. Note how the blue histograms for small galaxies peak at systematically larger b/a values, indicating that small galaxies (at fixed mass and redshift) are intrinsically rounder.}
\label{fig3}
\end{figure*}

\textbf{Fig. 3} compares histograms of $b/a$ of galaxies with different sizes in each redshift-mass bin. Since galaxies grow in size with time and mass, we cancel out this systematic effect by fitting the size-mass relation in each redshift-mass panel and use the residuals from these relations (i.e. $\Delta \log a$) as a measure of the size of the galaxies. For the fitting results see \citet{Fang2017}. The blue and red open histograms show the (normalized) $b/a$ of distributions galaxies in slices of small  $\Delta \log a$ and large $\Delta \log a$, respectively. The orange one shows distributions of all the galaxies in that redshift bin. We see first that the $b/a$ distributions of small and large galaxies are quite different in that the smaller objects (blue lines) tend to be clustered at larger $b/a$ values and therefore are intrinsically rounder, while the $b/a$ distributions of the larger galaxies (red lines) peak at a smaller value and have long tails, which implies that these objects are more likely to be prolate or oblate galaxies instead of spheroids. Thus if one simply models the single marginalized $b/a$ distribution, one may end up mistaking both small spheroidal galaxies and large prolate ones for oblate objects, since such a marginalization over the $\log a$ dimension tends to give  a flatter global $b/a$ distribution (which favors more oblate galaxies) than the ones where only the galaxies of a certain size are involved. To overcome this danger of modeling the marginalized $b/a$ distribution is the main motivation for this work, where the correlations between the size and the shape of galaxies are included simultaneously.


\subsection{Mock images from the VELA simulations} 
\label{sub:mock_images_from_the_vela_simulations}

Besides the data from observations, we also make use of the mock images generated from the VELA set of high-resolution hydrodynamic cosmological zoom-in galaxy simulations \citep[][and references cited there]{2014MNRAS.442.1545C,Ceverino:2015db,Zolotov:2015hk,Snyder2015,Tacchella:2016a,Tacchella:2016b}. The images are produced by SUNRISE, which is a parallel Monte Carlo code for the calculation of radiation transfer \citep{Jonsson:2006gt,JonssonPrimack:2010,Jonsson:2010fz}. In the generation of such mock images, emission lines, the effects of stellar evolution, scattering and absorption by dust, the resolution of the instrument (i.e., HST/WFC3), the PSF and sky background are included. We use \textsc{GALFIT} \citep{Peng:2010eh} to measure the structural parameters of all the J band (F125W) mock images of the galaxies with $1 < z < 2$ and the H band (F160W) images of the ones with $2 < z < 3$. By doing this we get $\sim6500$ good\footnote{A measurement is good when all the fitting flags defined by \citet{2012ApJS..203...24V} are zero.} \textsc{GALFIT} measurements for the images of 34 VELA galaxies from multiple orientation at a series of time steps. The reason why we choose this redshift range is that for most of the VELA galaxies, the simulation only runs down to $z=1$, while at redshifts higher than $z \sim 3$ the galaxies are too small (typically with $a\sim1\ \mathrm{kpc}$ at $3<z<4$) for \textsc{GALFIT} to measure the parameters robustly. The \textsc{GALFIT} setup is identical to the one used in \citet{2012ApJS..203...24V}. Note that for each galaxy at each time step, mock images viewed from 19 different directions are made, of which seven are viewed from random directions from one time step to the next. These images are qualitatively closest to the real observational data, and we call them the `truly random' cameras.


\section{Models} 
\label{sec:models}

\subsection{The model of individual galaxies} 
\label{ssub:the_model_of_individual_galaxies}

In our modeling of the two-dimensional $b/a-\log a$ distributions, a fundamental assumption is that a galaxy is modeled as a solid three-dimensional ellipsoid, with a set of intrinsic axes $\left(a, b, c\right)$, which satisfy $a \geq b \geq c$. The shape of a galaxy is completely determined by the two (intrinsic) axis ratios, $b/a$ and $c/a$. \textbf{Fig. 4} shows our definitions of prolate, oblate and spheroidal galaxies, which are identical to the ones used in \citet{vanderWel:2014ka} in order to directly compare the results. Note that, although technically defined as spheroidal, some galaxies can have quite small apparent $b/a$ because of an allowed intrinsic $c/a$ as small as 0.3. Thus actually we recommend that one rename the `spheroidal' galaxies defined by \textbf{Fig. 4} as `spheroidal or triaxial'. When a galaxy is observed from a certain direction, specified by the polar angle $\theta$ and the azimuthal angle $\phi$, the observed $b/a$ and semi-major axis $a$ can be calculated by measuring the shape and size of the projected two-dimensional ellipse. By projecting a galaxy randomly in the whole 4$\pi$ solid angular space, one gets the theoretical projected $b/a-\log a$ probability distribution of the galaxy, which will serve as one of the building blocks for our models. To differentiate this solid ellipsoid modeling with the S\'ersic model introduced in the Appendix, we call the former the `ellipsoidal modeling'.

\begin{figure}[htb]
\centering
\includegraphics[width=0.5\textwidth]{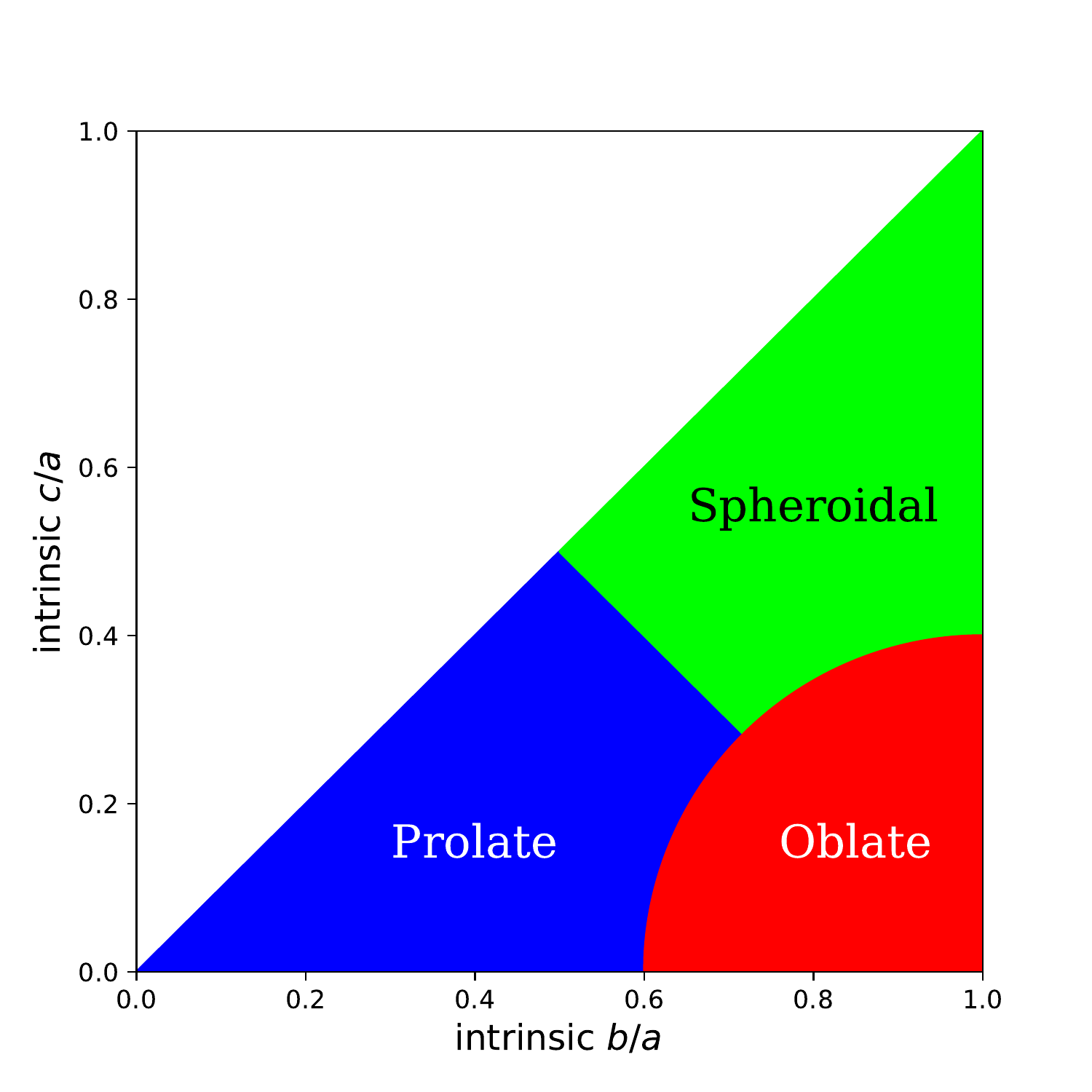}
\caption{The definition of the three shapes of galaxies, which is identical to that of \citet{vanderWel:2014ka} for the sake of direct comparison.}
\label{fig2}
\end{figure}

\textbf{Fig. 5} shows the juxtaposition of the $b/a-\log a$ distributions of six such galaxies generated by the S\'ersic modeling (left) and the ellipsoidal modeling (right), color-coded by the surface number density of points in their neighborhoods. Before we go into the features of different kinds of galaxies, we point out that in the ellipsoidal modeling, we have taken into account the measurement uncertainties in both $b/a$ and $\log a$ in the generation of these distributions by smearing them with an appropriate measurement uncertainty, which is adopted from \citet{2012ApJS..203...24V}. In principle, the ellipsoidal model may be criticized for not being realistic in the sense that it assumes the galaxies to have well-defined boundaries and are not transparent, which is not true. But from \textbf{Fig. 5} it is clear that the $b/a-\log a$ distributions generated by the ellipsoidal modeling and the more realistic S\'ersic modeling are similar in their shapes. Therefore we argue that the main conclusions in this work, which are made based on the ellipsoidal modeling, won't change qualitatively if we adopted this more realistic S\'ersic modeling. Given the different magnitudes of scatter of the distributions, the exact best fitting parameters and the fractions of different shapes may well change, though. For more discussion on the S\'ersic modeling see the Appendix.

For oblate galaxies (e.g., \textbf{Fig. 5 (a) and (b)}), the distributions are quite flat over a fairly large range of $b/a$. If a galaxy has perfectly round geometry when viewed face-on (i.e., intrinsic $b/a=1$), theoretically we would observe a flat distribution all the way up to apparent $b/a=1$. But due to the fact that random noise always causes the measured $b/a$ to be an underestimate of the real one \citep{Chang:2013bz}, a bump shows up at the relatively high $b/a$ end owing to the fact that the definition of $b$ and $a$ are inverted. Another reason for the existence of this bump is the imperfect oblateness, i.e., intrinsic $b < a$. In this case, the probability that the projected galaxy has an apparent $b/a$ that is close to the intrinsic one is larger, and thus there's a bump at the $\left(b/a\right)_{\mathrm{obs}} \sim \left(b/a\right)_{\mathrm{int}}$. On the other hand, due to the finite thickness of the disk, a second bump exists at the value around intrinsic $c/a$, which determines the value of the lowest projected $b/a$.

The distribution generated by a prolate galaxy is a curved trajectory \textbf{(e.g. Fig. 5 (c) and (d))}, and the shape of the curve depends on $\left(b/a, c/a\right)$. Another noteworthy feature is that a prolate galaxy is much more likely to be viewed edge-on, which induces small apparent $b/a$ and large $\log a$. We can see this by simply appreciating how large the difference is between the number densities in the upper left and lower right corners in \textbf{Fig. 5 (c)}.

A triaxial galaxy \textbf{(like Fig. 5 (e))} can be regarded as an intermediate phase between prolate and oblate objects, not only because the locus of such galaxies lies between the prolate and the oblate regions in \textbf{Fig. 4}, but also because a typical triaxial galaxy (e.g. \textbf{Fig. 5 (e)}) possesses features of both shapes: two bumps at lower and higher $b/a$ ends, which is characteristic of an oblate object, and a curved trend vs. $a$, which is a feature of prolate galaxies.

For a spheroidal galaxy \textbf{(Fig. 5 (f))}, the distribution is simply a small blob at large apparent $b/a$, with less-complex internal structure than oblate, prolate, or triaxial galaxies.

\begin{figure*}
\centering
\includegraphics[width=1.0\textwidth]{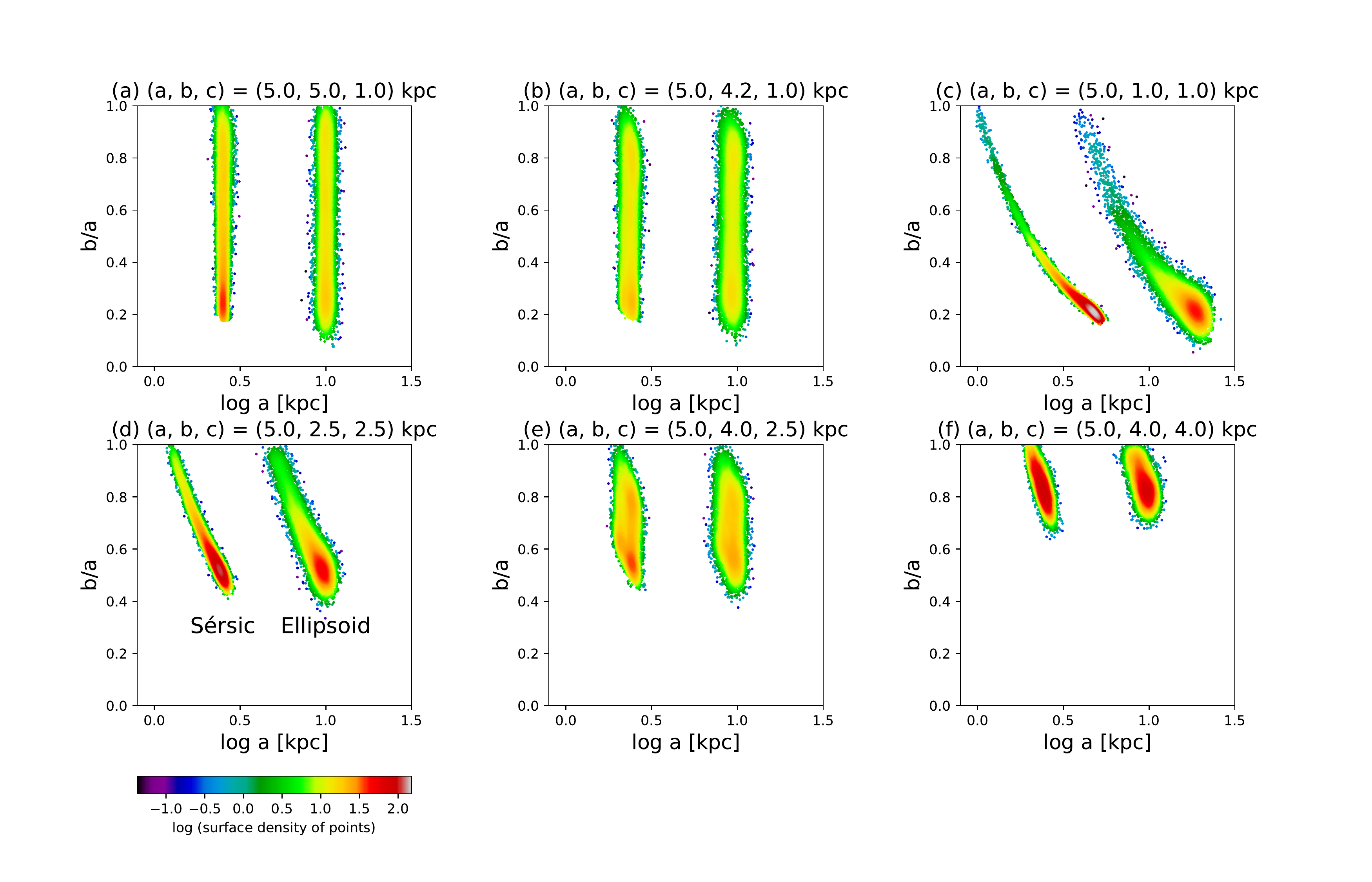}
\centering
\caption{Comparison between $b/a-\log a$ distributions of galaxies with different shapes, generated in the S\'ersic modeling, and those generated in the ellipsoidal modeling. The one in the left of each panel is from the S\'ersic modeling, and the other one in the right is from the ellipsoidal modeling. They are offset to the left and right for clarity. The points are color-coded by the number density of the points in their neighborhood. The distributions generated by the two models are qualitatively the same, although they have different magnitudes of scatters.} 
\end{figure*}

\subsection{Empirical modeling of the $b/a-\log a$ distributions} 
\label{sub:empirical_modeling_of_}

From the visual inspections of the $b/a-\log a$ distributions of all the different redshift-mass bins in \textbf{Fig. 1}, we can see some features smoothly evolving with time and mass. This implies the possibility of fitting the distributions of all the bins using those of the two most extreme bins, with $0.5 < z < 1.0$ and $10 < \log \left( M_{*} / M_\odot\right) < 10.5$ (circled by the red rectangle in \textbf{Fig. 1}, hereafter called the late-oblate bin),  and $2.0 < z < 2.5$ and $9 < \log \left( M_{*} / M_\odot\right) < 9.5$ (circled by the blue rectangle, hereafter called the early-prolate bin). We do such a simplified modeling to see whether it captures the picture, i.e., that the oblateness (prolateness) increases (decreases) with increasing time and mass. {\color{black}{Specifically, we assume that the model $b/a-\log a$ distribution $P(b/a, \log a)$ in any redshift and mass bin is a linear combination of those of the two extreme bins, $P_{\mathrm{prolate}}$ and $P_{\mathrm{oblate}}$, both of which are normalized: $P_{\mathrm{model}}(b/a, \log a ) = f \cdot P_{\mathrm{prolate}}(b/a, \log a ) + (1 - f) \cdot P_{\mathrm{oblate}}(b/a, \log a )$.
To calculate the best fitting weight of the early-prolate bin $f$, we renormalize $P_{\mathrm{model}}(b/a, \log a )$ so that it has the same normalization as the observed data, so that we can use the following formula to compute the log likelihood for the i-th $b/a-\log a$ bin: \citep{Holden:2012ef}

\begin{equation}
\log \ L_i = n_i\ \log(m_i) - m_i - \log(n_i !) \ ,
\end{equation}

where $n_i$ is the number of galaxies in this bin and $m_i$ is the number predicted by the model. Given a further assumption that different bins are independent of each other, the log of the total likelihood is simply a summation like
\begin{equation}
\log\ L = \sum_{i} \log\ L_i \ .
\end{equation}
By using the Markov Chain Monte-Carlo (MCMC) method we can find the best fitting weight of the early-prolate bin $f$.


\subsection{Fully quantative modeling of the $b/a-\log a$ distributions} 
\label{sub:fully_quantative_modeling_of_}


Based on the prior knowledge that a more realistic model should involve the correlation between the size and the shape of galaxies, we extend the model in \citet{vanderWel:2014ka} in this way: firstly we add a new dimension, $\log a$, and model the two-dimensional distribution of the data on the $b/a-\log a$ plane; secondly our model population has a multivariate normal distribution of $\left(E, T, \gamma=\mathrm{log\ }a\right)$, with mean $\left(\bar{E}, \bar{T}, \bar{\gamma}\right)$ and the covariance matrix $\Sigma$. Furthermore we allow only the covariance between $E$ and $\gamma$, i.e., $\mathrm{Cov}\langle E, \gamma \rangle$, to vary in the modeling, and both of the other two are set to zero. There are two reasons why we didn't involve more covariances: First the existence of the curved boundary in the data distribution indicates the change of the intrinsic $c/a$ with the size of a galaxy, which can be most directly attributed to the covariance between $E=1-c/a$ and $\gamma$; and second, we have found that adding more covariance doesn't help to get a better modeling of the data in the sense that the likelihood calculated based on Poisson statistics doesn't improve. To differentiate this purely mathematical model from the empirical one introduced later on, we call this model `the ETa model'.

We have also tried using a model in which we assume firstly that the $\left(a, b, c\right)$ of a galaxy can only be taken from a finite set ($\sim100$ sets of $\left(a, b, c\right)$), and secondly that the relative abundance of galaxies with different $\left(a, b, c\right)$ are independent and to be determined via the linear decomposition of the real $b/a-\log a$ distribution. We finally discard this model because the large number of free parameters leads to severe overfitting, which made the results meaningless.

Given a multivariate Gaussian distribution of $\left(E, T, \gamma\right)$, the probability that one observes a certain set of such parameters, $P\left( \left(E, T, \gamma\right) | \left( \bar{E}, \bar{T}, \bar{\gamma} , \Sigma\right) \right)$ can be easily calculated. And for each set of $\left(E, T, \gamma\right)$ we can calculate the probability distribution of its apparent $\left(b/a, \log a\right)$, i.e., $P\left(\left( b/a, \log a\right) | \left(E, T, \gamma\right) \right)$. Thus we can calculate the model $b/a-\log a$ distribution with the following formula:
\begin{equation}
\begin{aligned}
& P\left(\left(\frac{b}{a}, \log a\right) |\ \left( \bar{E}, \bar{T}, \bar{\gamma} , \Sigma\right)\right) = \\
&\sum P\left(\left(\frac{b}{a}, \log a\right) | \left(E, T, \gamma\right)\right) \cdot P \left( \left(E, T, \gamma\right) | \left( \bar{E}, \bar{T}, \bar{\gamma} , \Sigma \right)\right)
\end{aligned}
\end{equation}

By similar renormalization and MCMC to the ones described in Section 3.2, we can find the best fitting parameter set, which, in this case, is $\left(\bar{E}, \bar{T}, \bar{\gamma}, \Sigma\right)$.

\section{Results} 
\label{sec:results}

\subsection{Continuity of trends with redshift and mass} 
\label{sub:contibuity_of_trends_with_redshift_and_mass}

Before we show the quantitative results, we point out that there are some aspects evolving with time and mass that can be seen in \textbf{Figs. 1 and/or 3}. As the redshift decreases and the mass increases, we see the following trends:

Firstly, as is seen from \textbf{Fig. 3}, the $b/a$ distribution gets more and more uniform, and the peak of the number density at small $b/a$ gets less pronounced. For the late-oblate bin, the distribution is rather flat over a large $b/a$ range regardless of the size of a galaxy, which is characteristic of the growth of an oblate population.

Secondly, in each panel of \textbf{Fig. 1}, the upper right corner where few objects populate at low mass and high redshift gets filled up at higher mass and lower redshift. From the visual inspections on the images of the galaxies that are appearing here we find that these objects are predominantly disk galaxies, with well defined disks and bulges.

Also, the negative correlation between $A_{V}$ and $b/a$ in \textbf{Fig. 1} gets more and more pronounced. This is also consistent with the growth of an oblate population,  because only the oblate galaxies can have a larger path length through the whole galaxy (and therefore a larger $A_V$ value) and a smaller projected $b/a$ value simultaneously.

Finally, we can see that there's a small tail of the $b/a-\log a$ distributions (\textbf{Fig. 1}) at the lower right corner, and it gets more pronounced with increasing time and mass. This is understandable under the hypothesis that the oblateness grows with time and mass, because when an oblate galaxy is viewed in an edge-on configuration, it would possess a larger path length at the center, and a smaller one at the edge. Therefore the inner parts of the image will be more attenuated by dust than the outskirts, making the light intensity decline slower with the radius. Given the definition of the half-light radius, an oblate galaxy will therefore have a larger $a$ when viewed edge-on than it has when viewed face-on. The growing significance of this feature with time and mass is yet more evidence for a growing oblate population. The ellipsoidal model for edge-on oblate galaxies used here exhibits \emph{constant} semi-major axis as galaxies become more inclined and therefore does not model this extended tail to larger radii.  However, an empirical method to correct for the tail had little impact on the modeling results (see Appendix), and thus we believe that the tail is not a serious problem for this work. See the full comparison between the modeling results with and without correction in the Appendix.

These trends are all consistent with the big picture that the oblateness (prolateness) of galaxies increases (decreases) with time and mass, which is the major conclusion of this work. It also serves as a sanity check on our quantitative modeling results, in the sense that a reasonable modeling should give an increasing fraction of oblate objects with time and mass.

\subsection{Simplistic models don't work} 
\label{sub:simplistic_model}

Here we demonstrate that some simplistic models are not able to fit the observed data well, in order to motivate the complexity of the model ultimately adopted.  Two simple cases are illustrative: a pure oblate population and a pure prolate population.

For a model to fit the data well, it has to

\begin{itemize}
\item[(1)] recover the lower curved boundaries of $b/a-\log a$ distributions in all panels, at least qualitatively; and
\item[(2)] generate $b/a$ histograms of galaxies with different sizes that are at least qualitatively consistent with those shown in \textbf{Fig. 3}.
\end{itemize}

Bearing these requirements in mind, we argue that a pure prolate population is not sufficient to fit the data. Such a population is of course able to recover the lower curved boundaries, but by comparing \textbf{Fig. 5 (c)} and any panel from \textbf{Fig. 1} we can see that the number density difference between the lower right corner and the upper left corner of the diagram predicted by a prolate population is much larger than it is in the real data. Thus we conclude that even at the early-prolate bin, galaxies can't be all large and prolate. But they could be mixtures of large and smaller, more spheroidal galaxies.

Alternatively a pure oblate population with roughly constant intrinsic $c$ length regardless of intrinsic $a$ or $b$ can also generate a population with a curved boundary. This is shown in \textbf{Fig. 6 (a)}, which plots measured $b/a$ vs. $\log a$ for a series of mock galaxies with the same intrinsic $c$ length and the length of $a$ ranging from 1 to 10 kpc. The histograms for large, small, and all galaxies for the mock model are plotted in \textbf{Fig. 6(b)} for comparison to similar histograms in \textbf{Fig. 3}.  It is seen that the oblate model fails the second criterion above in the sense that the $b/a$ distribution of the large mock galaxies (red line) is quite flat, which is inconsistent with the distributions at $9 < \log\left(M_{*}/M_{\mathrm{\odot}}\right) < 10$ in \textbf{Fig. 3}, which tend to peak near $b/a \sim 0.3$.  The $b/a$ distribution of the small mock galaxies (blue line) is also wrong: it peaks at $b/a\sim0.4$ whereas the real small galaxies in \textbf{Fig. 3} peak at $b/a\sim0.7$, showing they are much rounder than the mock oblate model. Therefore we argue that a pure oblate population with constant intrinsic $c$ length is not able to explain the observed $b/a-\log a$ distribution, either.

\begin{figure}[htb]
\centering
\subfigure{\label{fig:subfig:a}\includegraphics[width=0.5\textwidth]{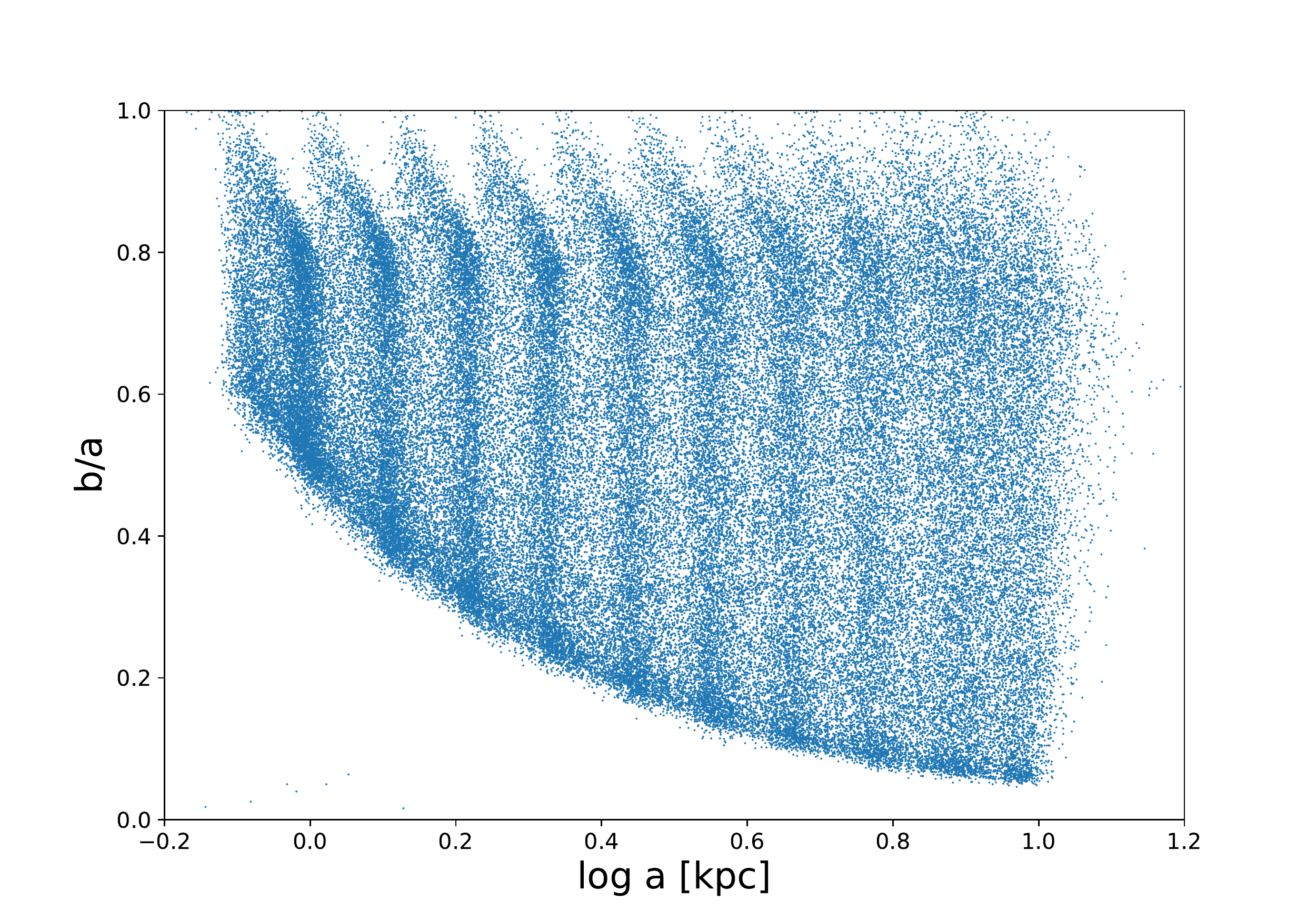}}
\subfigure{\label{fig:subfig:b}\includegraphics[width=0.5\textwidth]{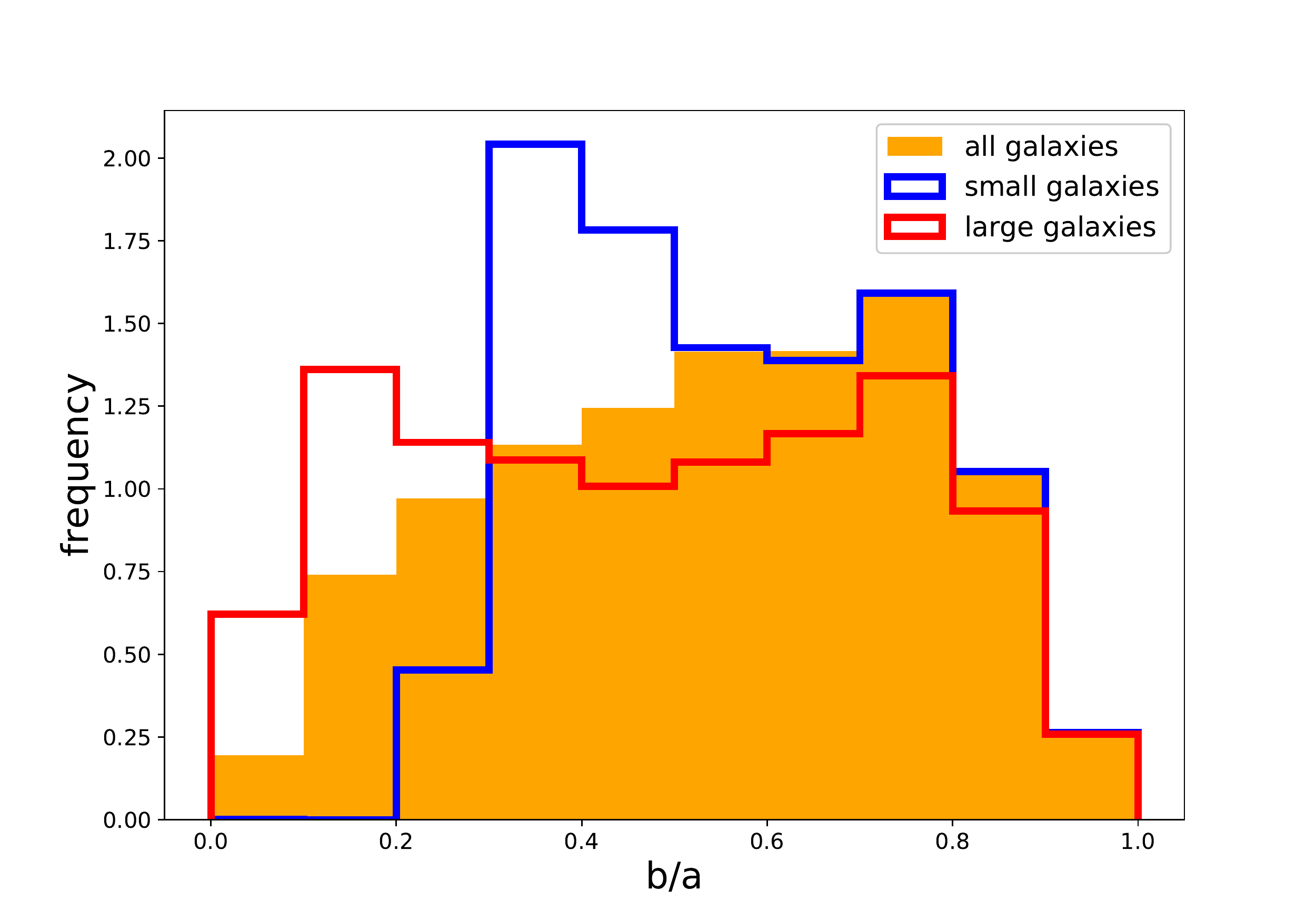}}
\caption{\textbf{Panel (a)}: the $b/a-\log a$ distribution of a series of oblate galaxies with the same intrinsic $c=0.5\ \mathrm{kpc}$, and intrinsic $b/a=0.8$. The existence of the vertical stripes is due to the fact that the intrinsic length $a$ of a galaxy was assumed to be one of ten discrete values. \textbf{Panel (b)}: The $b/a$ histograms of the mock galaxy populations with different sizes. Blue open histogram: small galaxies with $0.1 < \log a < 0.3$, with $a$ in kpc. Red open histogram: large galaxies with $0.7 < \log a < 0.9$. Orange filled histogram: all galaxies. Compared to similar histograms in \textbf{Fig. 3}, the red histogram (for large galaxies) is much flatter, and the blue histogram (for small galaxies) peaks at smaller values ($b/a\sim0.4$ here vs. $b/a\sim0.7$ in \textbf{Fig. 3}).}
\end{figure}


\subsection{Empirical modeling results} 
\label{sub:empirical_modeling_results}

\begin{figure*}
\includegraphics[width=1.0\textwidth]{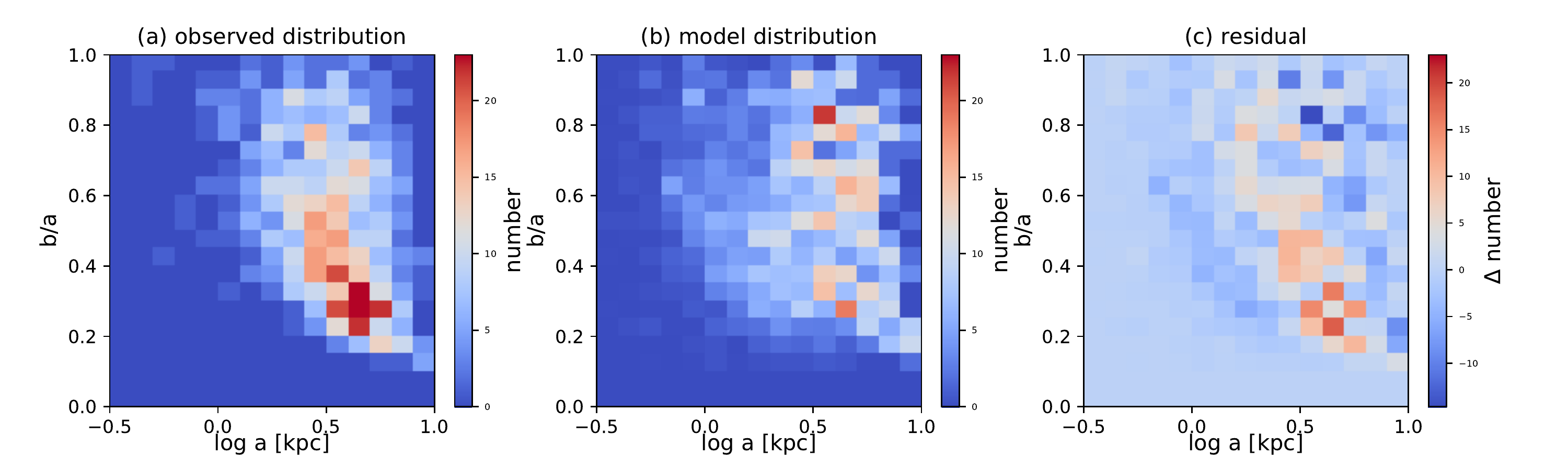}
\caption{The empirical modeling results of the galaxies with $0.5 < z < 1.0$ and $9.5 < \log \left( M_{*} / M_\odot\right) < 10$. \textbf{Panel (a)}: The observed $b/a-\log a$ distribution. \textbf{Panel (b)}: The model distribution. \textbf{Panel (c)}: The residual map. All bins are color-coded by the ($\Delta$) numbers of the galaxies in them. The model distribution is a linear combination of those of the early-prolate bin and the late-oblate bin, and renormalized so that the model has the same number of galaxies as the real observed data. A non-negligible systematic pattern is clearly seen in the residual map, which indicates that the empirical model is not a perfect fit.}
\end{figure*}

\begin{figure}[htb]
\includegraphics[width=0.5\textwidth]{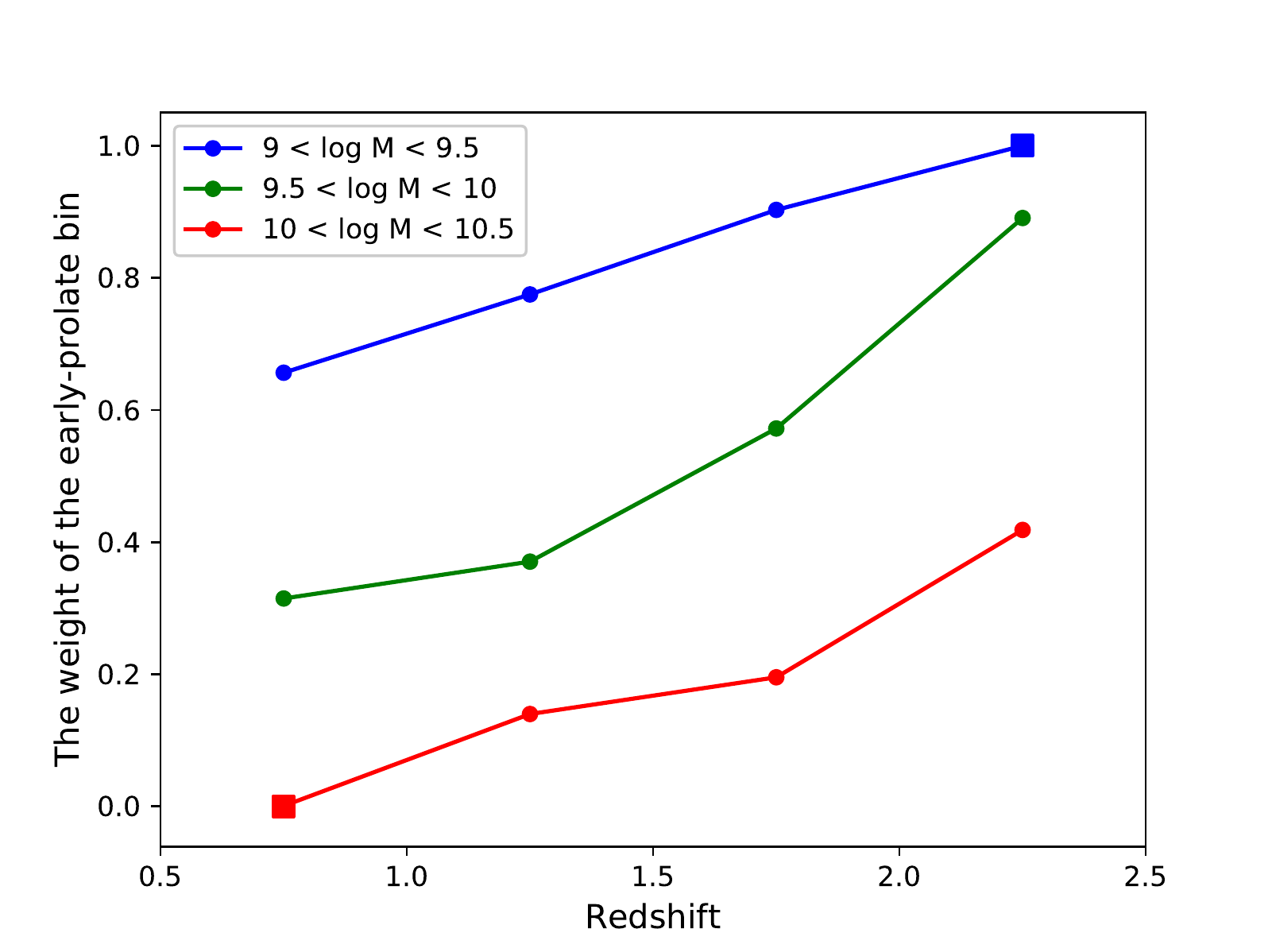}
\caption{The evolving trend of the relative weight of the early-prolate bin) with time and mass. The squares denote the cases in which the early-prolate or late-oblate bins are themselves modeled, so the weight of the early-prolate bin is naturally unity or zero.}
\end{figure}

We turn now to the results of more complex models, starting with the empirical model described in Section 3.2, which fits all $b/a-\log a$ distributions as combinations of low-mass high-z galaxies (early-prolate bin) and high-mass low-z galaxies (late-oblate bin).  \textbf{Fig. 7} shows the results of this fit for one mass-redshift panel ($0.5 < z < 1.0$ and $9.5 < \log \left( M_{*} / M_\odot\right) < 10$), including the observed data, the best fitting model and the residual map. \textbf{Fig. 8} shows the evolving trends of the relative weight $f$ of the early-prolate bin in the decomposition with time and mass given by this empirical modeling. As we can see, the importance of the prolate galaxies decreases with increasing time and mass. \textbf{Fig. 9} summarizes the fractions of the three shapes of galaxies in all the redshift and mass bins given by the empirical modeling.  To obtain these values, we assume the fractions in a certain redshift and mass bin are linear combinations of those in the early-prolate bin and the late-oblate bin, weighted by the same weights as the $b/a-\log a$ distribution itself, while the fractions of the two extreme bins are obtained from ETa modeling.} From \textbf{Figs. 8 and 9} it appears that the oblateness (prolateness) increases (decreases) with increasing time and mass, which is consistent with the continuous evolution discussed in Section 4.1.

Nevertheless, we do see some systematic patterns in the residual maps of such empirical fittings (e.g. the residual map of \textbf{Fig. 7}), which is not surprising because, even though the $b/a-\log a$ distributions evolve smoothly, it is unreasonable to expect that all such mass-redshift bins can be fit perfectly with two extrema.

\begin{figure*}
\includegraphics[width=1.0\textwidth]{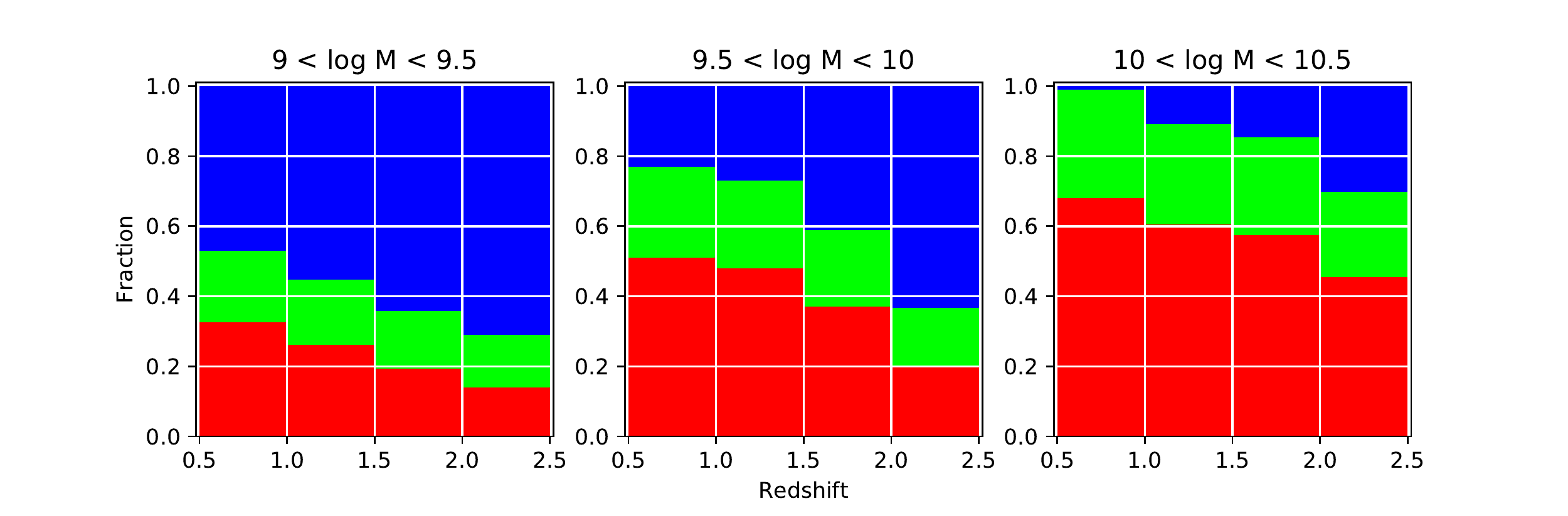}
\caption{The evolution of the fractions of different shapes of the star-forming galaxies in CANDELS with redshift and stellar mass, given by the empirical modeling. Blue bars: The fractions of prolate galaxies. Green bars: The fractions of spheroidal galaxies. Red bars: The fractions of oblate galaxies. Compare with the ETa modeling results in \textbf{Fig. 14}.}
\end{figure*}


\subsection{ETa modeling results} 
\label{sub:quantitative_modeling_results}

Here we present the results of  the ETa modeling for all the redshift-mass bins. To be concise, \textbf{Figs. 10 and 11} show relevant plots (i.e. the observed distribution, the model distribution, the residual map, and the model distributions of galaxies of the three shapes defined by \textbf{Fig. 4}) only for the two extrema among all the bins, namely the late-oblate bin and the early-prolate bin\footnote{Similar plots illustrating the rest of redshift-mass bins can be found at \url{https://sites.google.com/site/zhw11387/Home/research}}. 

\begin{figure*}
\centering
\includegraphics[width=1.0\textwidth]{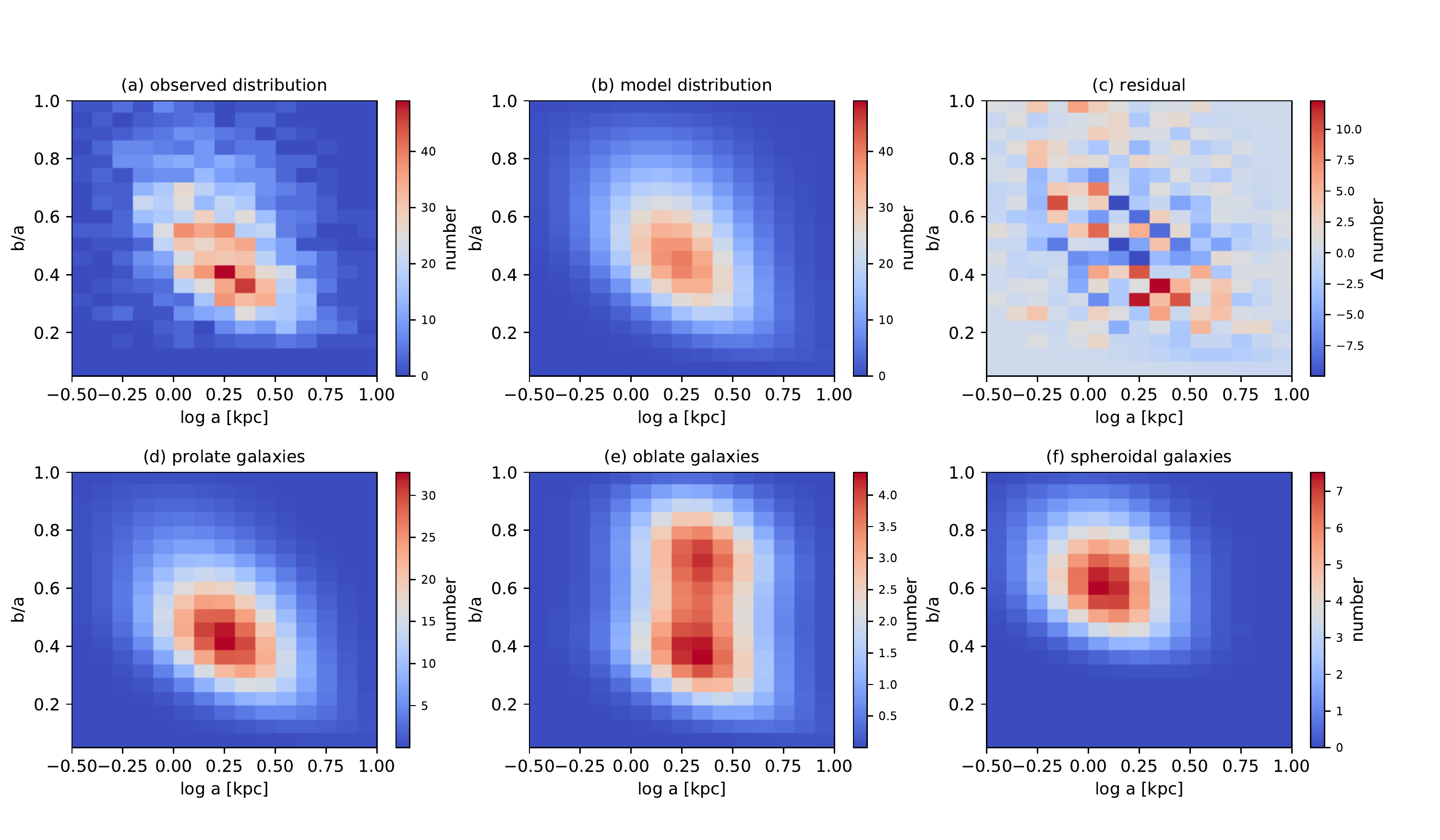}
\caption{Illustrative plots of the fitting results for the $b/a-\log a$ distribution of galaxies in the early-prolate bin. \textbf{Panel (a)}: observed data distribution. \textbf{Panel (b)}: best fitting model distribution. \textbf{Panel (c)}: residual map. \textbf{Panel (d)}: The distribution of prolate galaxies. \textbf{Panel (e)}: The distribution of oblate galaxies. \textbf{Panel (f)}: The distribution of spheroidal galaxies. Note the smaller numbers of oblate and spheroid galaxies compared with prolate ones.}
\end{figure*}

\begin{figure*}
\centering
\includegraphics[width=1.0\textwidth]{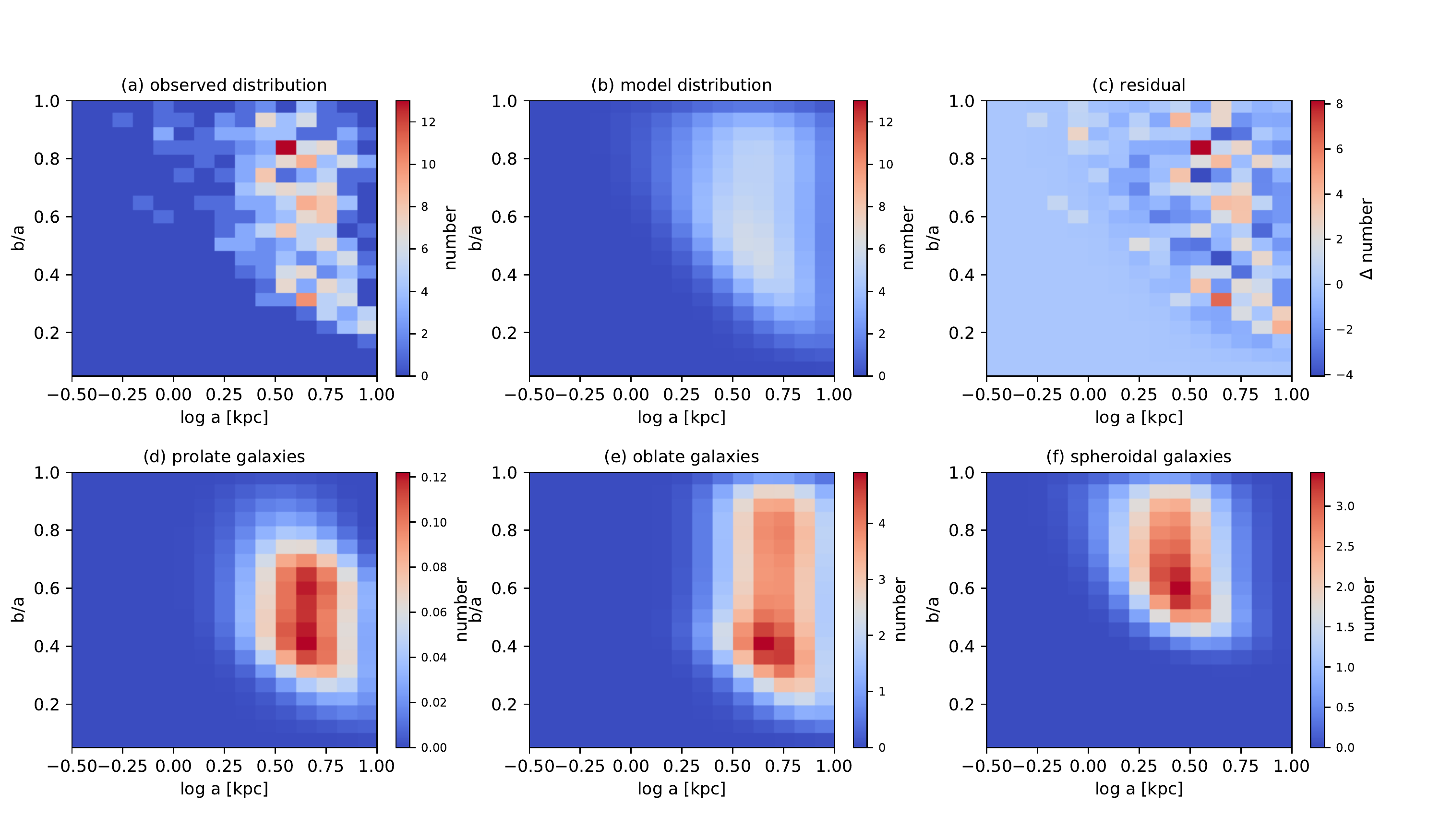}
\caption{The same plots as the previous figure, but for the galaxies in the late-oblate bin. Note that the bin has the smallest number of galaxies at any bin.}
\end{figure*}

From \textbf{Figs. 10 and 11} we can immediately see that the early-prolate and late-oblate bins are indeed dominated by prolate and oblate galaxies, respectively. All the $b/a-\log a$ bins are color-coded by the number of (either real or model) galaxies in them. In the early-prolate bin, most of the galaxies are prolate, especially in the lower right corner of the projected $b/a-\log a$ diagram, while in the late-oblate bin, we can barely find any prolate objects. The best fitting parameters and the fractions of the three galaxy shapes of each redshift-mass bin are tabulated in Table 1. For the most massive bins, there are too few galaxies for the bootstrap algorithm to get a realistic estimation of the uncertainties of the parameters; thus we only present the best fitting parameters values without errors.  Qualitatively the fraction of prolate (oblate) galaxies in the prolate (oblate) bin is consistent with what \citet{vanderWel:2014ka} found. But our results differ from the previous work in the sense that we find more prolate and/or spheroidal galaxies and fewer oblate disks than \citet{vanderWel:2014ka}, especially in the low redshift and low mass bins. This is expected because modeling the marginalized projected $b/a$ distribution is likely to mistake large prolate galaxies and small round galaxies for oblate objects, given the fact that the smaller galaxies tend to be rounder. To illustrate this point we further present \textbf{Fig. 12}, panel (a) of which shows the $b/a$ distributions of the larger, smaller and all the galaxies with $0.5 < z < 1.0$ and $9.5 < \log \left( M_{*} / M_\odot\right) < 10$, and panel (b) shows the $b/a$ distribution of all the galaxies analyzed by \citet{vanderWel:2014ka} in that redshift-mass bin.  It is clear seen from \textbf{Fig. 12 (a)} that the smaller galaxies have a broad and somewhat flat distribution of $b/a$ at $b/a \gtrsim 0.4$ (blue open histrogram), which is indicative of a spheroidal population (maybe marginally oblate) according to the definitions shown in \textbf{Fig. 4}. On the other hand, the $b/a$ distribution of the larger galaxies (red open histogram) shows a significant peak at $b/a\sim 0.25$, and declines as $b/a$ increases. This is strongly characteristic of a prolate population with a intrinsic $c/a\sim 0.25$. However, by marginalizing over the $\log a$ dimension (i.e. superposing the red and the blue histogram), we get a relatively broad and flat $b/a$ distribution over a large $b/a$ distribution, which is very similar to the distribution shown in \textbf{Fig. 12 (b)} that was analyzed by \citet{vanderWel:2014ka}. Therefore we argue that \citet{vanderWel:2014ka} end up misidentifying small objects with median $b/a$ and large ones with $b/a\sim 0.25$ as oblate disks. From this example we can see that adding the dimension of galaxy size helps to avoid such misidentifications, which is one of the main points of this work. Our modeling results in \textbf{Figs. 10 and 11} also match our intuition that spheroidal galaxies are predominantly smaller than prolate and oblate galaxies.

\begin{figure}[htb]
\centering
\subfigure{\label{fig:subfig:a}\includegraphics[width=0.5\textwidth]{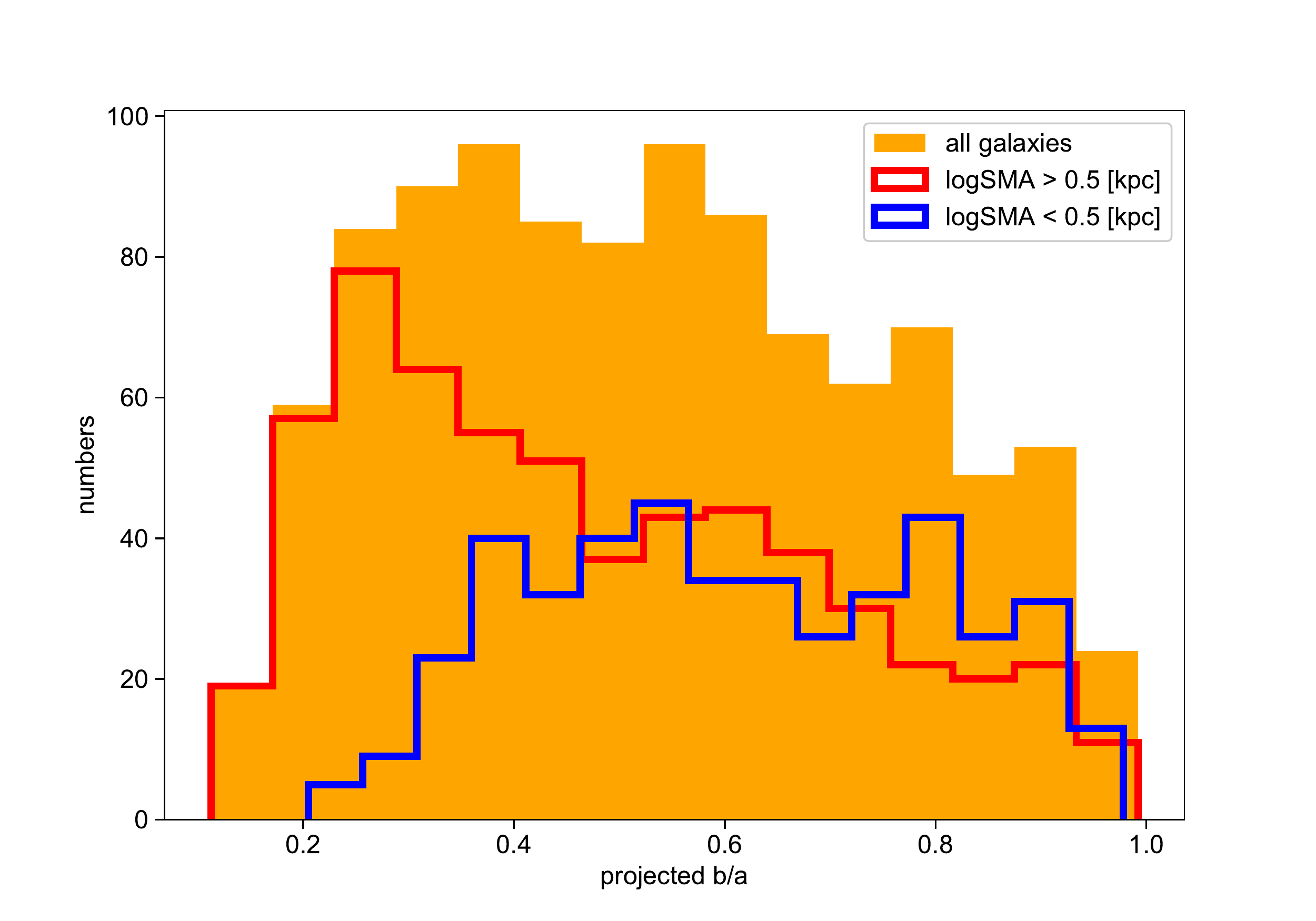}}
\subfigure{\label{fig:subfig:a}\includegraphics[width=0.5\textwidth]{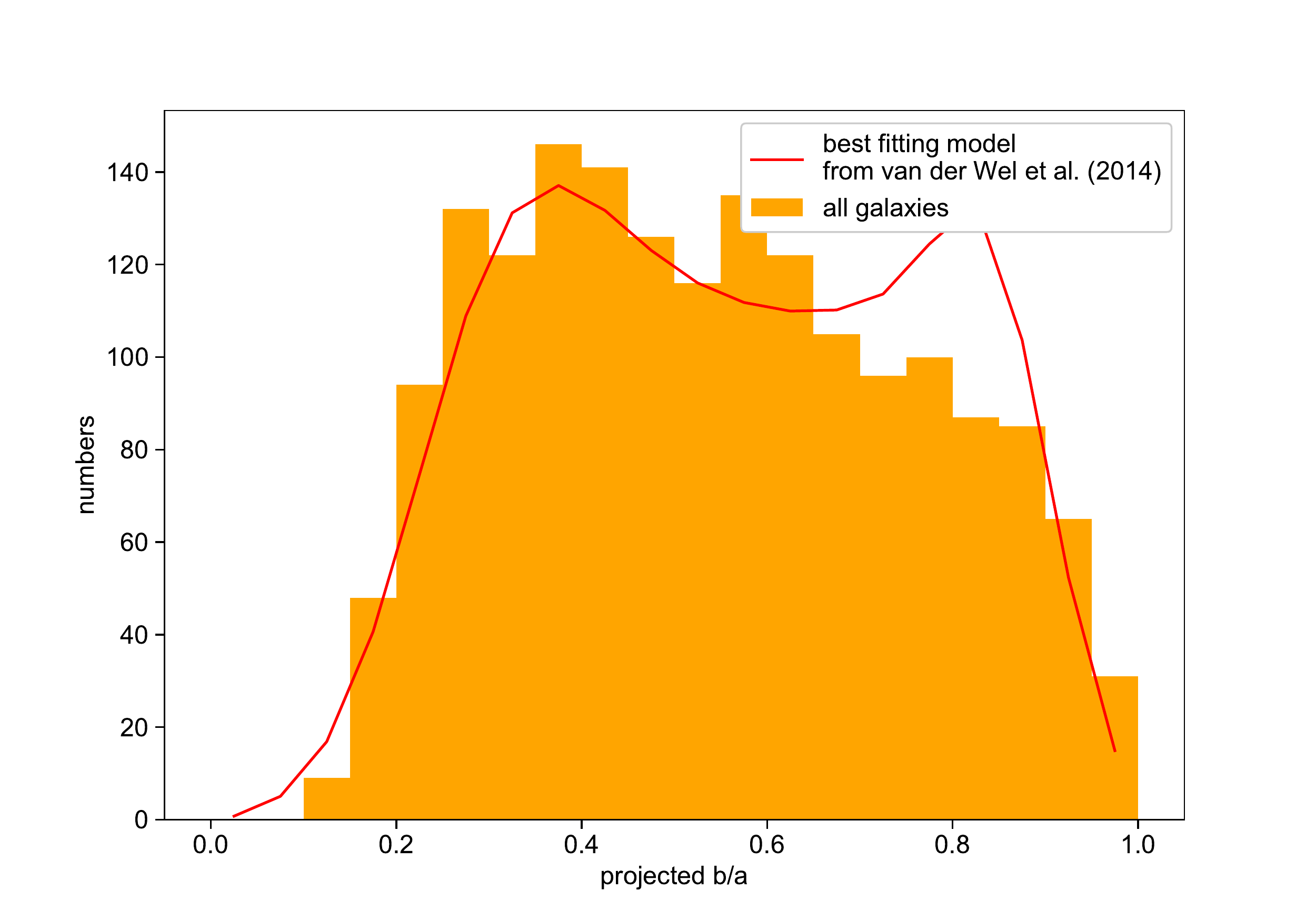}}
\caption{\textbf{Panel (a)}: The $b/a$ distribution of the CANDELS galaxies with $0.5 < z < 1.0$ and $9.5 < \log \left( M_{*} / M_\odot\right) < 10$. Red open histogram: $b/a$ distribution of larger galaxies with $\log a > 0.5$. Blue open histogram: the same distribution of smaller galaxies with $\log a < 0.5$. Orange filled histogram: the same distribution of all the galaxies in that redshift-mass bin.
 \textbf{Panel (b)}: The $b/a$ distribution of the 3D-HST galaxies with $0.5 < z < 1.0$ and $9.5 < \log \left( M_{*} / M_\odot\right) < 10$ (the one that was analyzed by \citet{vanderWel:2014ka}). Orange filled histogram: $b/a$ distribution of all the galaxies in that redshift-mass bin. Red Curve: the best fitting model obtained by \citet{vanderWel:2014ka}}
\end{figure}

\begin{figure*}
\centering
\includegraphics[width=0.95\textwidth]{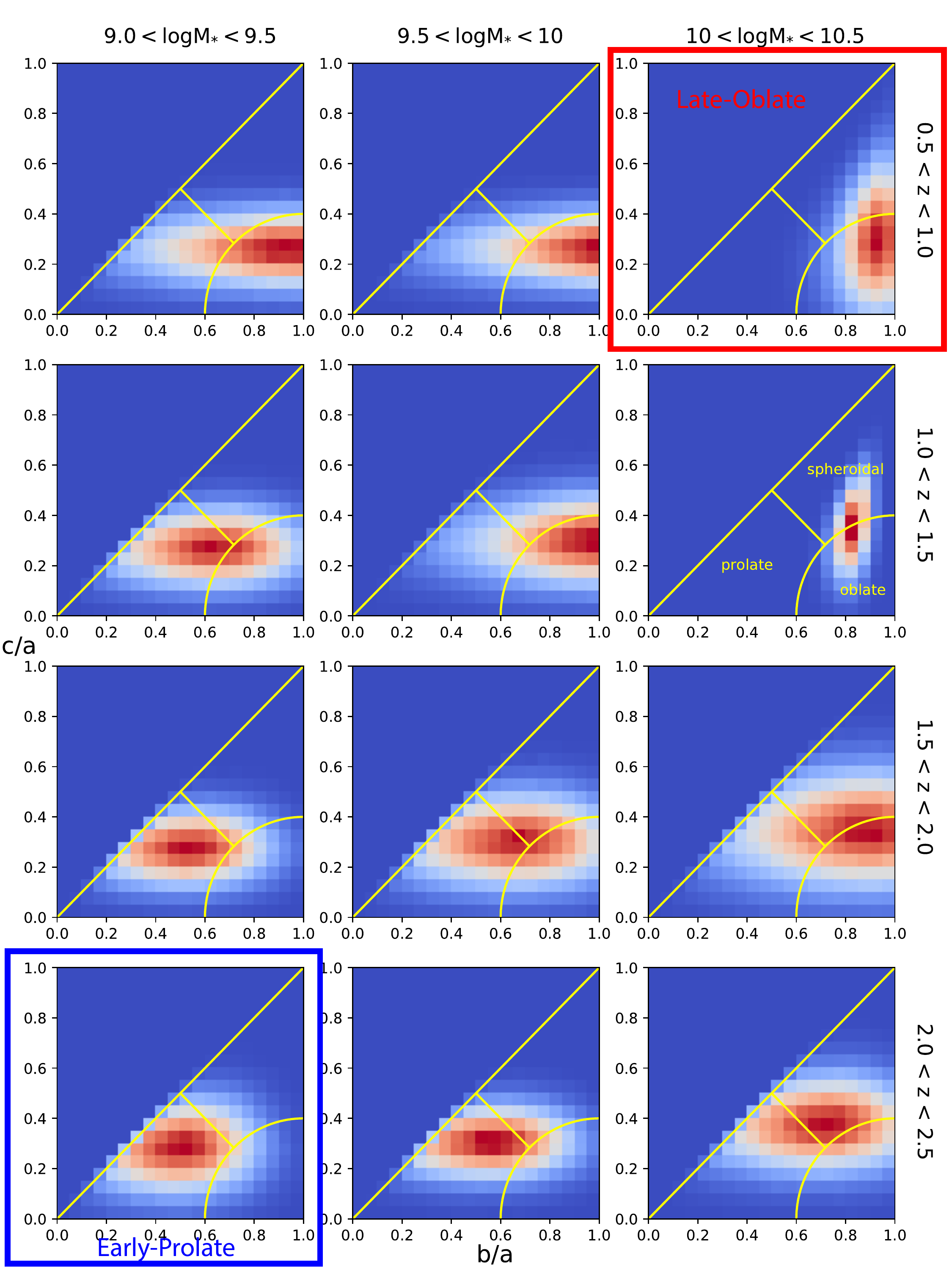}
\caption{The evolution of the $b/a-c/a$ distributions with redshift and mass. The distributions are generated by the best fitting parameters listed in Table 1. The yellow lines are the boundaries between different galaxy shapes as defined in \textbf{Fig. 4}. The redder color means larger densities of the distribution. Note the evolving trend with increasing time and mass that the peak of the distributions gradually move from the prolate region to the oblate/spheroidal region.}
\end{figure*}

\textbf{Fig. 13} shows the evolution with redshift and mass of the $b/a-c/a$ distributions generated by the best fitting parameters listed in Table 1. The yellow lines are the boundaries between different galaxy shapes as defined in \textbf{Fig. 4}. Again we can see that the peak of the $b/a$ distributions generally moves from the prolate region to the oblate/spheroidal region, which is another confirmation of the picture that the oblateness (prolateness) increases (decreases) with increasing time and mass.

\textbf{Fig. 14} shows the evolution trends of the three fractions with time and stellar mass. Again we can see a general trend that the fraction of prolate (oblate) galaxies decreases (increases) with time and mass, consistent with the picture that oblate disks emerge and come to dominate the whole galaxy population with the cosmic time and the stellar mass of galaxies, which is in good agreement with the trend seen in \textbf{Fig. 9}.  By comparing this plot with fig. 4 of \citet{vanderWel:2014ka}, we further confirm that we find more prolate and/or spheroidal galaxies than they did.

\begin{figure*}
\includegraphics[width=1.0\textwidth]{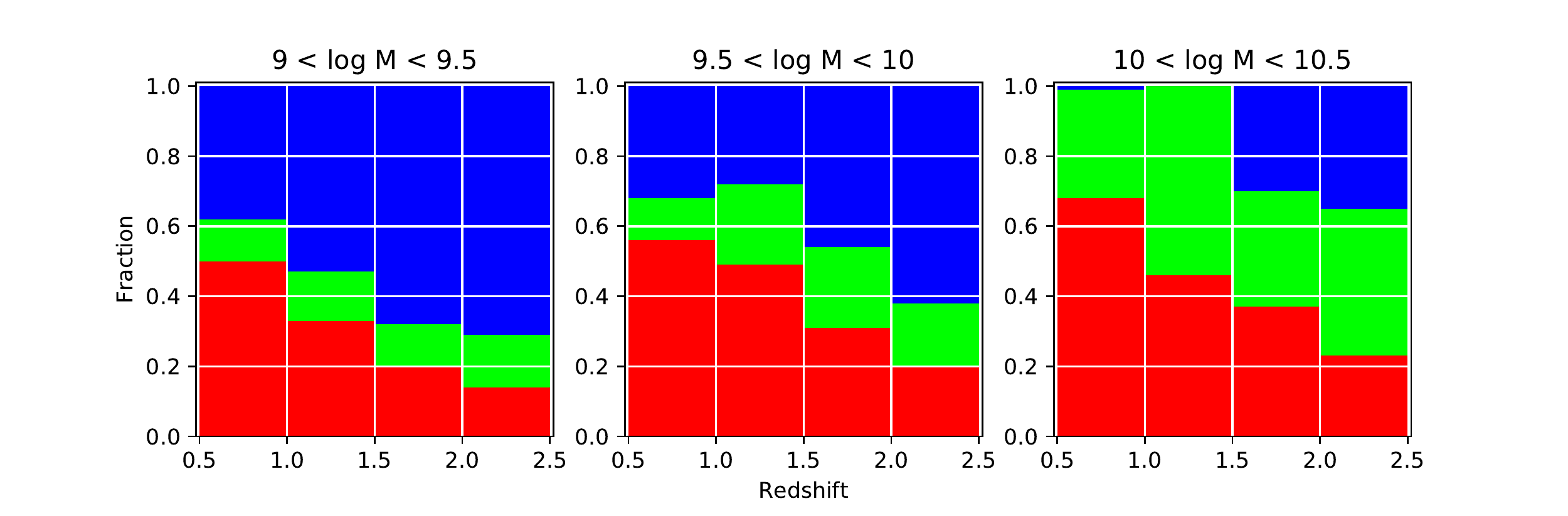}
\caption{The evolution of the fractions of different shapes of the star-forming galaxies in CANDELS with redshift and stellar mass, given by the ETa modeling. Blue bars: The fractions of prolate galaxies. Green bars: The fractions of spheroidal galaxies. Red bars: The fractions of oblate galaxies. These fractions are qualitatively in good agreement with those obtained by the empirical modeling \textbf{in Fig. 9.}}
\end{figure*}

\begin{sidewaystable*}
\centering
\vspace{0.7\textwidth}

\begin{tabular}{ccccccccccccc}
\hline
\hline
Redshift & $\mathrm{log}\left(M_{*}/M_\odot\right)$ & $\bar{E}$\tnote{a} & $\bar{T}$\tnote{b} & $\bar{\gamma}=\langle\log a\rangle$\tnote{c} & $\mathrm{Var}\left(E\right)$\tnote{d} & $\mathrm{Var}\left(T\right)$ & $\mathrm{Var}\left(\gamma\right)$ & $\mathrm{Cov}\langle E,\ \gamma\rangle$ & $f_{\mathrm{prolate}}$ & $f_{\mathrm{oblate}}$\tnote{e} & $f_{\mathrm{spheroidal}}$\tnote{e} & $N_{\mathrm{galaxy}}$\\
\hline
$0.75$ & $9.25 $ & $0.747\pm0.005$ & $0.49\pm0.07$ & $0.462\pm0.005$    &   $0.009\pm0.002$ &  $0.77\pm0.04$ & $0.046\pm0.002$  & $0.018\pm0.002$ & $0.376\pm0.008$ & $0.51\pm0.01$ & $0.12\pm0.01$ & 2071  \\
$0.75$ & $9.75 $ & $0.74\pm0.01$ & $0.16\pm0.08$ & $0.57\pm0.01$     &   $0.008\pm0.006$ &  $0.66\pm0.07$ & $0.043\pm0.003$  & $0.016\pm0.005$ & $0.324\pm0.009$ & $0.56\pm0.03$ & $0.12\pm0.03$ & 1024 \\
$0.75$ & $10.25$ & $0.728$ & $0.166$ & $0.680$    &   $0.035$ &  $0.039$ & $0.062$  & $0.038$ & $0.013$ & $0.678$ & $0.309$ & 426  \\
$1.25$ & $9.25$ & $0.740\pm0.003$ & $0.85\pm0.01$  & $0.436\pm0.003$    &   $0.009\pm0.002$ &  $0.21\pm0.02$ & $0.047\pm0.002$  & $0.017\pm0.001$ & $0.53\pm0.01$ & $0.33\pm0.02$ & $0.13\pm0.01$ & 2531  \\
$1.25$ & $9.75$ & $0.711\pm0.004$ & $0.2\pm0.1$   & $0.527\pm0.007$    &   $0.012\pm0.005$ &  $0.4\pm0.1$ & $0.048\pm0.002$  & $0.020\pm0.003$ & $0.28\pm0.02$ & $0.49\pm0.04$ & $0.23\pm0.02$ & 1319  \\
$1.25$ & $10.25$ & $0.659$ & $0.362$ & $0.604$    &   $0.016$ &  $0.007$ & $0.050$  & $0.023$ & $0.003$ & $0.541$ & $0.456$ & 525  \\
$1.75$ & $9.25$ & $0.736\pm0.004$ & $0.978\pm0.004$  & $0.397\pm0.004$     &   $0.009\pm0.002$ &  $0.13\pm0.01$ & $0.048\pm0.002$  & $0.014\pm0.002$ & $0.68\pm0.01$ & $0.20\pm0.02$ & $0.118\pm0.009$ & 2714  \\
$1.75$ & $9.75$ & $0.710\pm0.004$ & $0.93\pm0.01$ & $0.496\pm0.005$     &   $0.013\pm0.003$ &  $0.32\pm0.03$ & $0.050\pm0.003$  & $0.020\pm0.002$ & $0.31\pm0.01$ & $0.46\pm0.02$ & $0.23\pm0.01$ & 1349\\
$1.75$ & $10.25$ & $0.673$ & $0.570$ & $0.619$     &   $0.020$ &  $0.577$ & $0.057$  & $0.027$ & $0.295$ & $0.338$ & $0.367$ & 464\\
$2.25$ & $9.25$ & $0.74\pm0.03$ & $0.98\pm0.02$  & $0.313\pm0.008$     &   $0.01\pm0.01$ &  $0.10\pm0.05$ & $0.052\pm0.003$  & $0.014\pm0.004$ & $0.71\pm0.06$ & $0.14\pm0.09$ & $0.15\pm0.03$ & 2099  \\
$2.25$ & $9.75$ & $0.71\pm0.02$ & $0.96\pm0.01$ & $0.40\pm0.01$     &   $0.008\pm0.006$ &  $0.15\pm0.04$ & $0.041\pm0.002$  & $0.011\pm0.002$ & $0.62\pm0.05$ & $0.20\pm0.06$ & $0.18\pm0.02$ & 1239\\ 
$2.25$ & $10.25$ & $0.642$ & $0.802$ & $0.492$     &   $0.014$ &  $0.295$ & $0.050$  & $0.018$ & $0.348$ & $0.229$ & $0.423$ & 550\\
\hline
\hline

\end{tabular}
\caption{Best fitting model parameters and fractions of the three shapes of each redshift-mass bin.}
\begin{tablenotes}
\footnotesize
\item{(a) }{$E=1-c/a$ is the ellipticity of a galaxy.}
\item{(b) }{$T=(a^2-b^2)/(a^2-c^2)$ is the triaxiality of a galaxy.}
\item{(c) }{$\gamma=\log a$.}
\item{(d) }{$\mathrm{Var}\left(E\right)$ is the variance of the Gaussian distribution of $E$.}
\item{(e) }{The definitions of the three shapes are from the boundaries in \textbf{Fig. 4}.}
\item{    }{For the $10 < \log \left( M_{*} / M_\odot\right) < 10.5$ bins, the bootstrap algorithm give unrealistic uncertainties due to the relative smaller numbers of observed galaxies in these bins, thus we only present the best fitting parameter values without errors.}
\end{tablenotes}
\end{sidewaystable*}


\subsection{Massive galaxies with $10 < \log \left(M_{*} / M_\odot\right) < 10.5$} 
\label{sub:massive_galaxies}

As can be seen from \textbf{Fig. 1}, the numbers of galaxies in the most massive bins are too small to be modeled robustly. But some qualitative comments can still be made.

Firstly we can clearly see that the curved boundary gets less and less pronounced as time goes by. There are two compatible explanations to this phenomenon: It can be either due to the fact that the number of prolate galaxies decreases with time and the oblate disks get more and more prevalent, so that the curved boundary, which is a natural outcome of a dominant prolate population, gradually fades out and a flat distribution over a large $b/a$ range takes its place. Alternatively, it can be due to the disappearance of the small and round star-forming galaxies. They start to quench and consequently drop out of the star-forming sample with time, leaving the upper left corner of the $b/a-\log a$ diagram less and less populated. See \citet{2013ApJ...765..104B} for more details of such a process.

The second trend to be seen in \textbf{Fig. 1} is that the correlation between $A_V$ values and $b/a$ improves with time and mass. At high redshift, there's barely a systematic trend of $A_V$ value with $b/a$, while as we move to lower redshift bins, the correlation gets more and more significant. At the late-oblate bin, the negative correlation between $A_V$ and $b/a$ is the most pronounced. This evolution of the negative correlation is consistent with the picture that the oblateness grows with time and mass and that the high-$A_V$ objects are disky edge-on galaxies with large dust path-lengths at late cosmic times.  This hypothesis is explored further in Section 5.3.


\section{Discussion} 
\label{sec:discussion}

\subsection{Comparison with VELA simulation images} 
\label{sub:comparison_with_vela_simulation_images}

We've pointed out that our results show growing oblateness of star-forming galaxies with time and stellar mass, which is also seen in the VELA simulations. However, \citet{Ceverino:2015db} and \citet{Tomassetti:2015ed} investigated the evolution of the three-dimensional {\it mass} profile in the VELA galaxies, while what we've modeled are the distributions of $b/a$ and $\log a$ measured from the {\it light} profiles of the galaxies. Thus a more direct comparison between the light profiles of CANDELS and VELA galaxies is necessary.

\begin{figure}[htb]
\centering

\subfigure[CANDELS galaxy]{\label{fig:subfig:a}\includegraphics[width=0.2\textwidth]{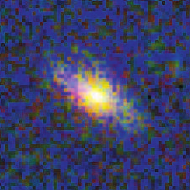}}
\hspace{0.01\linewidth}
\subfigure[VELA galaxy]{\label{fig:subfig:b}\includegraphics[width=0.2\textwidth]{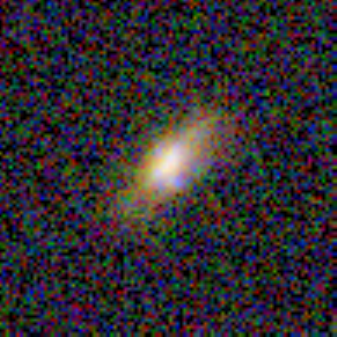}}

\caption{\textbf{Panel (a)}: an example of a large and elongated galaxy in CANDELS. This galaxy has a $z=2.27$ and $\log \left(M_{*}/M_\odot\right)=9.82$. \textbf{Panel (b)}: image of the simulated galaxy VELA05 at $z=1.32$, which has a prolate three-dimensional mass profile shape, including the effects of stellar evolution, dust scattering and absorption, the HST/WFC3 PSF, and sky background. Despite the bulge$+$disk appearance of the VELA galaxy, it is in fact prolate, showing that true 3D shapes cannot be reliably measured from projected images alone. The CANDELS galaxy, with similar appearance, is a member of a mass-redshift bin where most galaxies are modeled as prolate.} 
\label{fig3}
\end{figure}

To support the argument that our claimed detection of prolate galaxies in most mass and redshift bins is real (mostly at the lower right corner of the $b/a-\log a$ diagram), we have inspected numerous multi-waveband images of CANDELS galaxies in that corner and those of VELA prolate galaxies. \textbf{Fig. 15} shows two such images. \textbf{Fig. 15 (a)} is a typical galaxy located at the lower right corner of the $b/a-\log a$ diagram from CANDELS, and \textbf{Fig. 15 (b)} is a galaxy in its prolate phase at redshift $z=1.32$ in the VELA05 simulation. We can see that the two galaxy images share some common features: a brighter centroid with symmetric and extended linear structure. Given the similarity between the morphologies of these two galaxies, it's quite plausible that many galaxies in the lower right corner are indeed prolate, and pure visual inspection of the images is not sufficient to tell prolate objects from disks because they can be very similar in their projected light profiles.

\begin{figure}[htb]
\centering

\subfigure[CANDELS galaxies]{\label{fig:subfig:a}\includegraphics[width=0.5\textwidth]{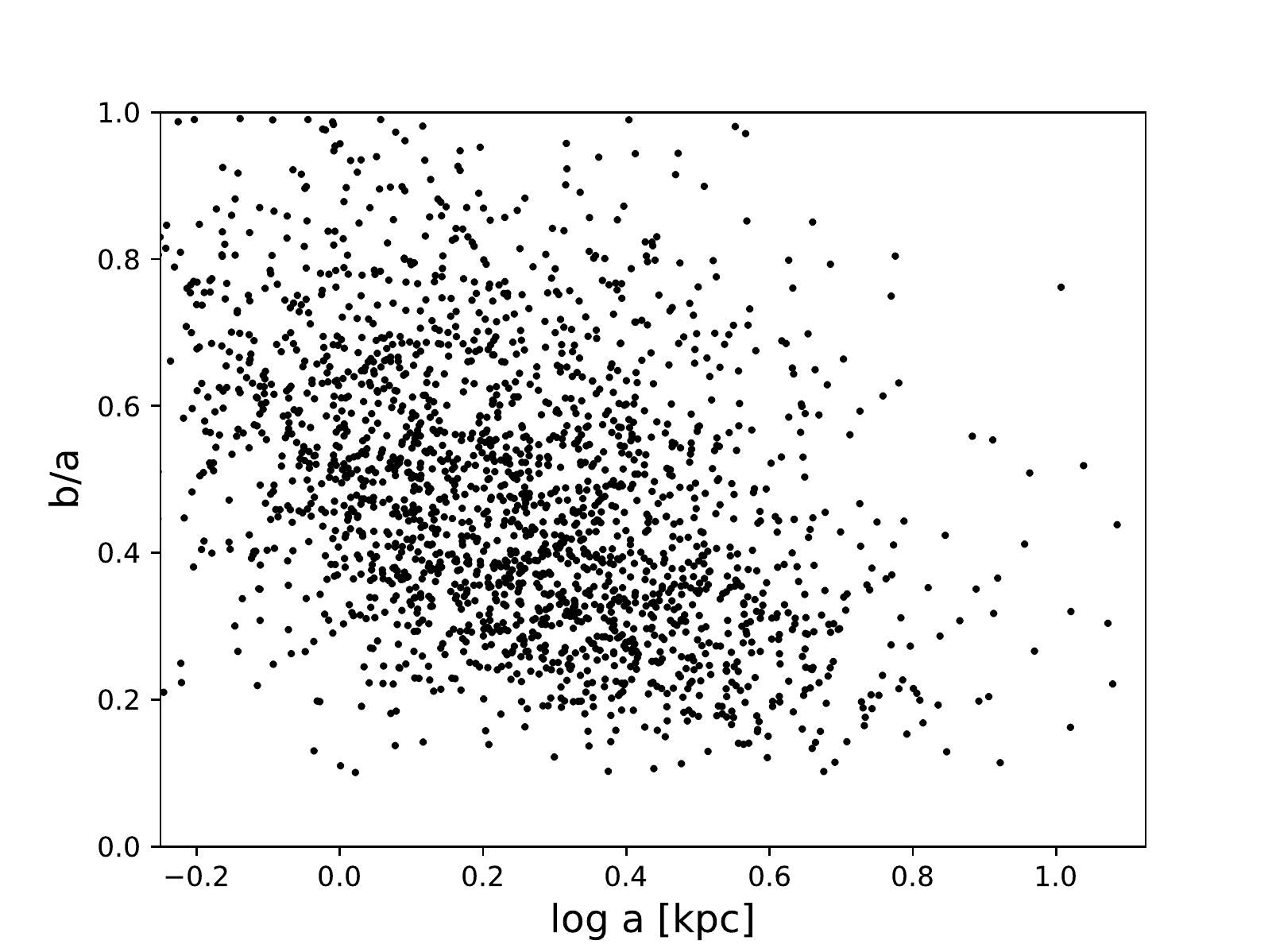}}
\vspace{0.01\linewidth}
\subfigure[VELA galaxies]{\label{fig:subfig:b}\includegraphics[width=0.5\textwidth]{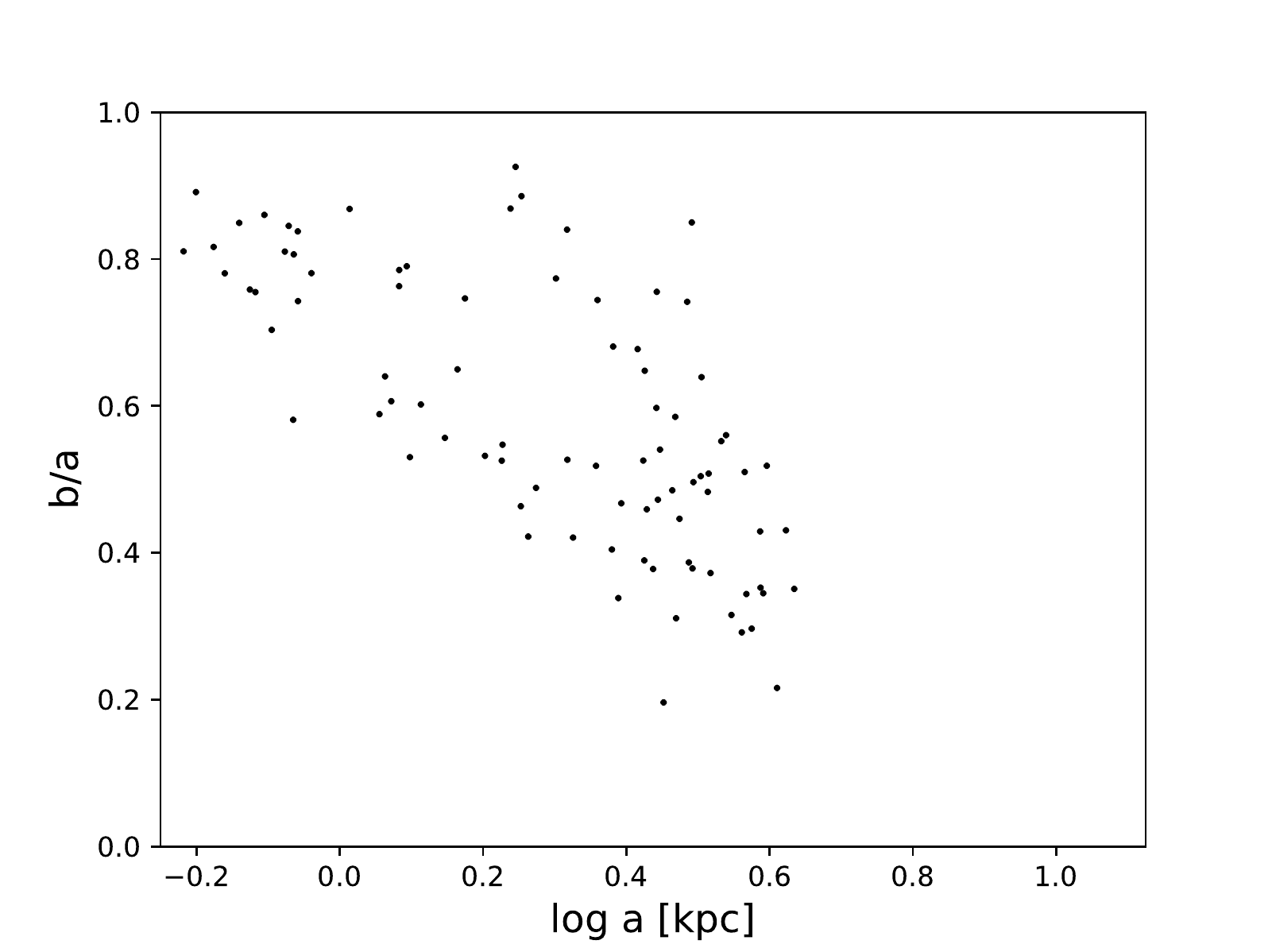}}

\caption{\textbf{Panel (a)}: the $b/a-\log a$ distribution of CANDELS star-forming galaxies in the early-prolate bin. \textbf{Panel (b)}: The same distribution of prolate VELA galaxies in all mass-redshift bins.  Prolate galaxies are defined as having a three-dimensional mass profile shape obtained by \citet{Tomassetti:2015ed} that places them in the prolate region defined by \textbf{Fig. 4}. Only the truly random images as defined in Section 2.3 are plotted. There is an overall agreement between the model and observed distributions, but the VELA galaxies are on systematically rounder and larger in size than the CANDELS galaxies.} 
\label{fig14}
\end{figure}

Besides totally qualitative comparisons based on the visual inspections on the images, we also compare the $b/a-\log a$ distributions of the galaxies from CANDELS and those whose mass profile shapes are in the prolate region of \textbf{Fig. 4} from the VELA simulation. \textbf{Fig. 16 (a)} shows the distribution of CANDELS galaxies in the early-prolate bin, and \textbf{Fig. 16 (b)} shows that of the prolate galaxies in the VELA simulation. In this panel we define as prolate objects with a three-dimensional mass profile shape obtained by \citet{Tomassetti:2015ed} that places them in the prolate region as defined by \textbf{Fig. 4}. We included all the prolate galaxies with $1 < z < 3$ viewed from truly random camera directions in \textbf{Fig. 16 (b)}, regardless of their masses, because the VELA simulations include rather few simulated galaxies (only 34 galaxies in the simulations, six of which are included in \textbf{Fig. 16 (b)}).  The statistics about the $\log a$ and $b/a$ of these two samples are included in the following \textbf{Table 2}:

\begin{table}
\begin{tabular}{ccccc}
\hline
\hline
Sample & $\langle \log a \rangle$ & $\sigma_{\log a}$ & $\langle b/a \rangle$ & $\sigma_{b/a}$ \\
CANDELS & 0.24 & 0.26 & 0.48 & 0.19\\
VELA & 0.29 & 0.25 & 0.59 & 0.18\\
\hline

\end{tabular}
\caption{Statistics about the $\log a$ and $b/a$ of the galaxies in the two samples in Fig. 16.}
\end{table}

From both the visual inspection on Fig. 16 and the statistics shown in Table 2 we can see that the $b/a-\log a$ distributions of these two  samples are consistent with each other, if we take into account the small number of points in the VELA sample. Specifically, they share the following features: 

\begin{itemize}
\item[(1)]clear curved lower boundaries from the small and round region ($-0.2 < \log a < 0,\ b/a > 0.6$) to the large and elongated region ($0.6 < \log a < 1.0,\ 0.2 < b/a < 0.4$);
\item[(2)]low frequency of objects at the upper right corner.
\end{itemize}
All of these are consistent with a population dominated by prolate galaxies. Therefore this serves as further supporting evidence that our modeling results are plausible. We also note that the VELA galaxies seems slightly rounder than the CANDELS galaxies, which can be seen from both the comparison between the two panels of Fig. 16, and the comparison between the lowest $b/a$ boundaries of Fig. 1 and the absence of galaxies with $c/a < 0.2$ in Fig. 17. We suggest that this is the result of the small number statistics of the VELA galaxies, given the fact that they're merely 34 galaxies observed from different directions at a series of timesteps, and that this small quantitative difference doesn't change our main qualitative conclusions.

\begin{figure}[htb]
\includegraphics[width=0.5\textwidth]{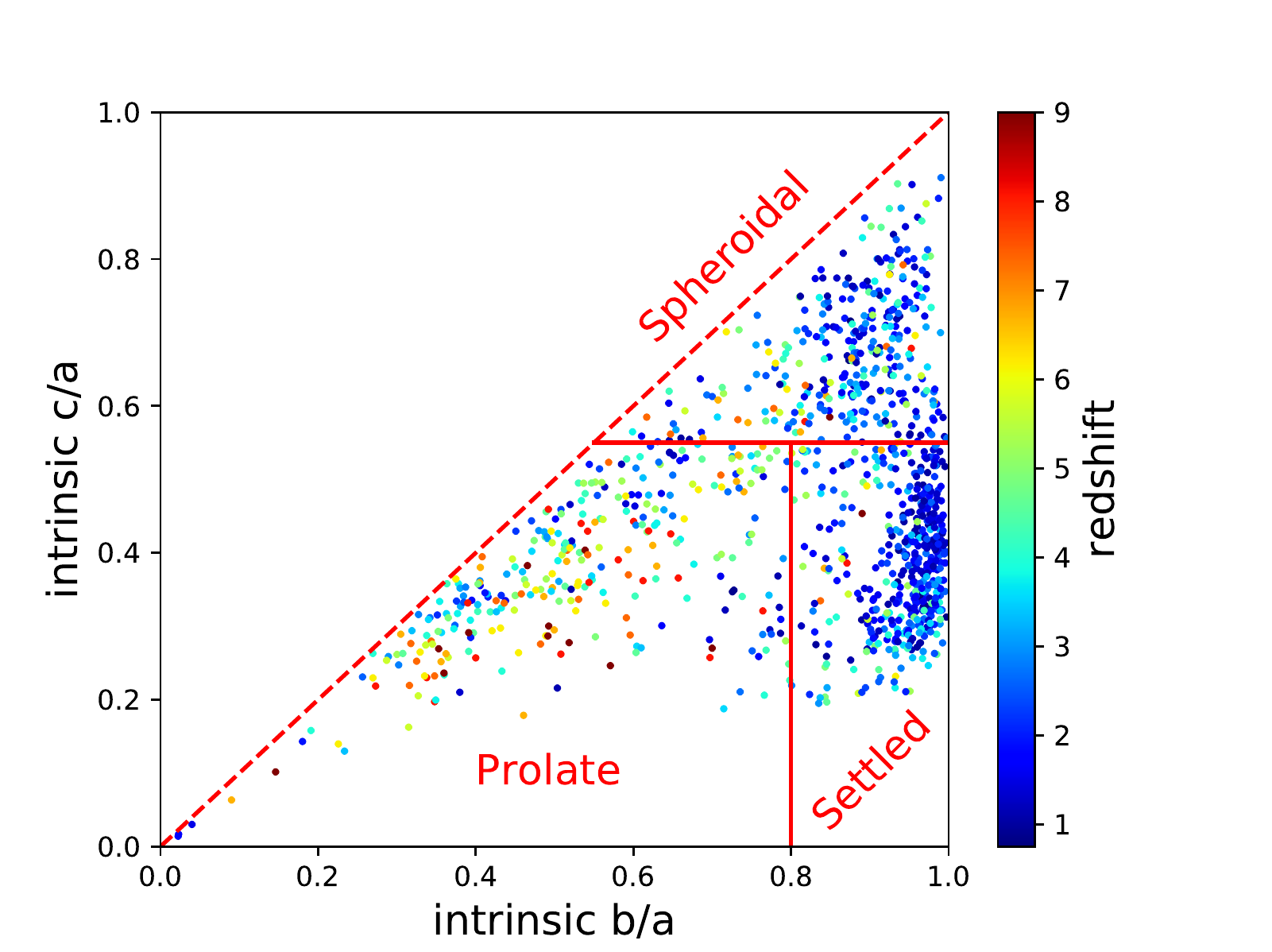}
\caption{The $c/a-b/a$ distribution of the VELA galaxies throughout the simulation. Each dot denotes the mass profile of a galaxy at a certain time step, which was obtained by \citet{Tomassetti:2015ed}, color-coded by its redshift. The red solid lines are the boundaries between different shapes of galaxies. The red dashed line is the physical boundary where $c=b$.}
\end{figure}

A third comparison between the CANDELS data and the VELA simulation data can be done in terms of the time evolution of the fractions of different shapes of galaxy stellar mass distributions. Before going into the comparison, we point out that many VELA galaxies don't reach a phase where they have a very small intrinsic $c/a$, and also that in the $c/a-b/a$ parameter space there's a cluster of galaxies right at the boundary between the oblate and spheroidal galaxies as defined by \textbf{Fig. 4}. We therefore argue that in order to do a fair comparison, figuring out new definitions of the shape of the galaxies is necessary. \textbf{Fig. 17} shows the distribution of the mass profiles of the VELA galaxies in $c/a-b/a$ parameter space, along with new boundaries between different shapes. Each point corresponds to a galaxy at a time step, color-coded by its redshift.

\begin{figure*}
\centering
\subfigure[]{\label{fig:subfig:a}\includegraphics[width=0.45\textwidth]{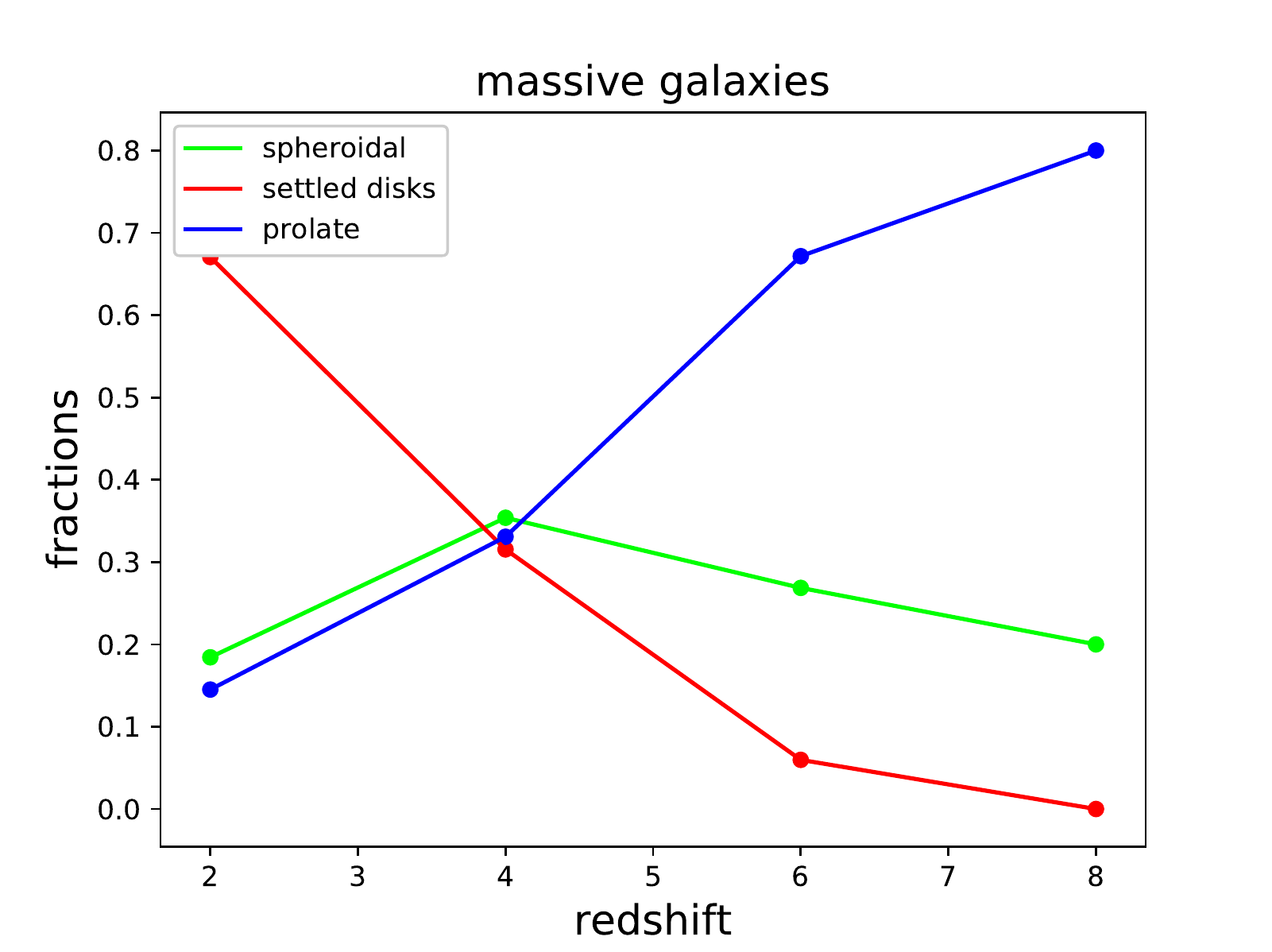}}
\hspace{0.01\linewidth}
\subfigure[]{\label{fig:subfig:b}\includegraphics[width=0.45\textwidth]{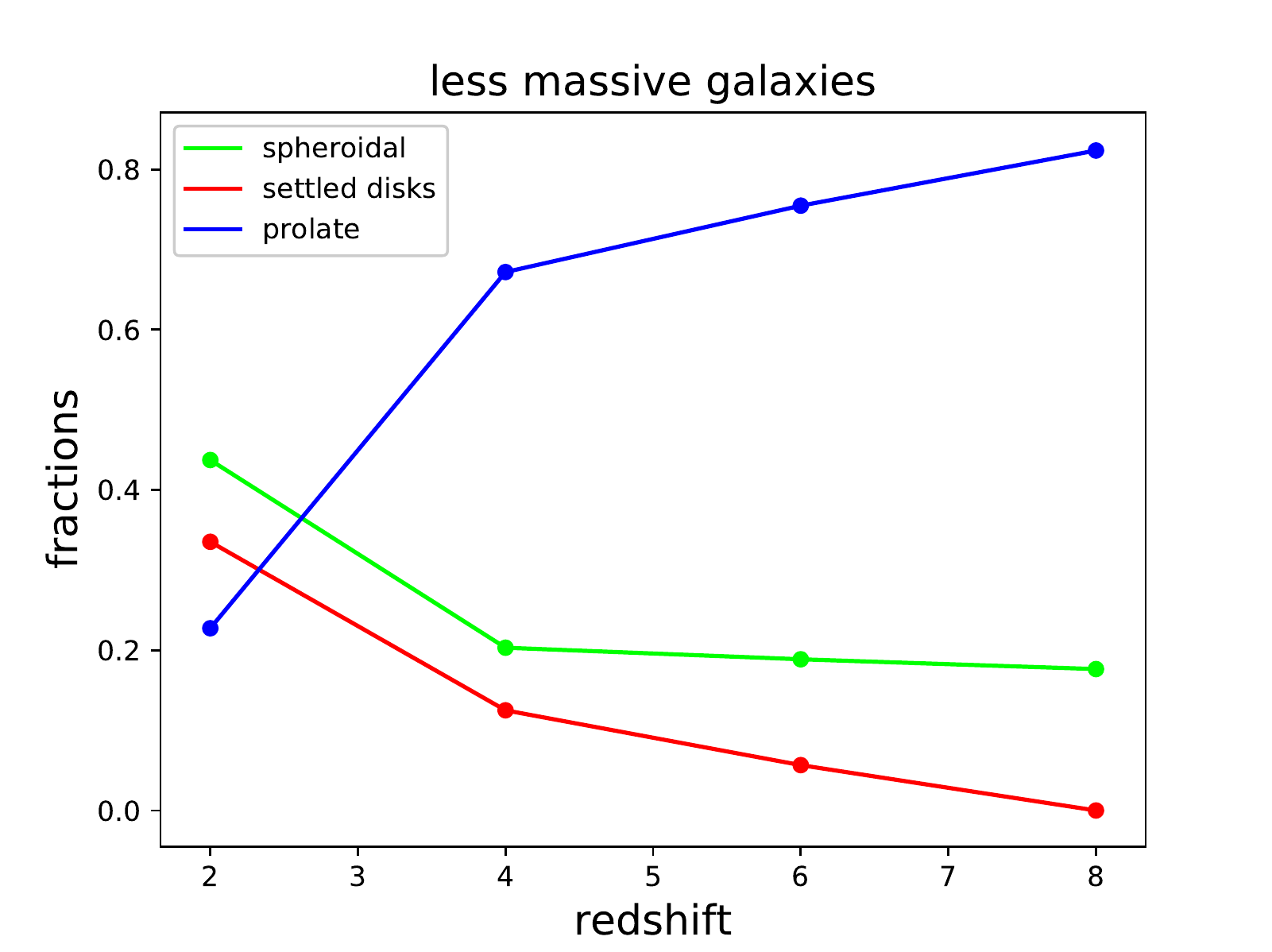}}
\caption{\textbf{Panel (a)}: The time evolution of the fractions of three different shapes of the VELA galaxies as defined in \textbf{Fig. 17}, with only the more massive galaxies  included. \textbf{Panel (b)}: The same evolution but for the less massive VELA galaxies.} 
\end{figure*}

To study the effect of stellar mass on the shape evolution we need to find criteria differentiating between larger and smaller VELA galaxies. We split the whole VELA sample into more massive galaxies and less massive ones using a critical stellar mass at a certain time step (in the VELA simulations, each time step is labeled with the scale factor of the universe $a_{\mathrm{exp}}$). By visually inspection on the $M_{*}-a_{\mathrm{exp}}$ diagram, we choose $M_{*}=6\times10^9M_{\mathrm{\odot}}$ at $a_{\mathrm{exp}}=0.26$ ($z=2.85$) as the critical mass. Galaxies with $M_{*}$ below (above) this value at $a_{\mathrm{exp}}=0.26$ are called low-mass (high-mass) galaxies.

Based on these definitions, we can now investigate the time evolution of the fractions of different shapes of the galaxies in different mass bins. \textbf{Fig. 18} shows two such evolutions for low- and high-mass galaxies. From both panels a clear trend is seen that the fractions of the prolate (settled) galaxies decrease (increase) with time. Furthermore we can see that at a given redshift, the fraction of the prolate objects among the more massive galaxies is smaller than it is among the less massive ones. In other words, the less massive galaxies are evolving like the more massive ones, but their evolution is relatively delayed in time. These trends, again, are consistent with the modeling of CANDELS $b/a - \log a$ distributions above (both the ETa and the empirical models), with the finding by \citet{Jiang2018} that the intermediate isophotes of more massive galaxies at lower redshift are more disky, with previous analysis on the evolution of three-dimensional shapes of the VELA galaxies by \citet[][see fig. 2]{Ceverino:2015db} and \citet[][see fig. 7]{Tomassetti:2015ed}, with kinematic observations of real galaxies by \citet{Kassin:2012kz,Simons:2016bk,Simons:2017bc}, and with kinematic calculations of simulated galaxies by \citet{Kassin:2014jp} and \citet{Ceverino:2017im} that found an increasing rotational support with time in star-forming galaxies.

Finally, we note that in the VELA simulations there is a gradual transition from prolate to oblate that tends to occur when the galaxy is in the vicinity of a characteristic stellar mass, $10^{9.5-10.0} M_{\mathrm{\odot}}$, and this transition mass does not show a significant variation with redshift \citep{Tomassetti:2015ed}. Typical masses in the same range are associated with the major wet compaction into a blue-nugget phase that triggers central quenching, and causes transitions in most structural, kinematic and compositional galaxy properties \citep{Zolotov:2015hk,Tacchella:2016a,Tacchella:2016b}.  In the CANDELS data we detect a similar mass dependence of the shape, transitioning from prolate to oblate in roughly the same mass range, but with the transition mass varying with redshift, with the scale crudely decreasing from $\sim 10^{10} M_{\mathrm{\odot}}$ at $z=2.5$ to $\sim 10^{9.3} M_{\mathrm{\odot}}$ at $z=1$ (\textbf{Fig. 9}). This may be a real difference between the simulations and observations, and if so it is worth putting forward as an interesting challenge for theoretical understanding. However, in principle it could also be due to systematics in the shape analysis and/or in the selection, either in the simulations or in the observations. For example, a transition-mass range that is decreasing with time may appear in the simulated sample as constant with time because the masses are monotonically increasing with time.

\subsection{The probabilities of a galaxy being prolate, oblate or spheroidal} 
\label{sub:the_probabilities_of_being_prolate_oblate_or_spheroidal_of_a_galaxy}

Given the best fitting model, we can easily calculate the numbers of prolate, oblate, or spheroidal galaxies in an arbitrary $b/a-\log a$ bin of CANDELS. If we further divide these numbers by the total numbers of galaxies in this bin we can get the probabilities of having a certain class of shape at a given set of projected $\left(b/a,\log a\right)$. Since the $b/a-\log a$ distribution evolves with time and mass, this modeling effectively enables us to predict how likely a galaxy is to be intrinsically prolate, oblate, or spheroidal \textit{as a function of its redshift, stellar mass, projected $b/a$ and $\log a$}. \textbf{Figs. 19} and \textbf{20} show such probability distributions for the star-forming galaxies in the high-z low-mass (early-prolate) and low-z high-mass (late-oblate) bins}, respectively\footnote{Such probability maps illustrating the rest of the redshift and mass bins can be found at \url{https://sites.google.com/site/zhw11387/Home/research}.}. As can be seen in both plots, the probability of being oblate at large $\log a$ and $b/a$ is always high (typically with a value of $0.7-1.0$), while the probability of being prolate at the lower right corner depends on time and mass. For the early-prolate bin, due to the dominance of the prolate population, the probability is high at this corner, since prolate galaxies are much more likely to show up at this region; while in the late-oblate bin, our modeling finds barely any prolate galaxies, which results in a high probability of being oblate for a galaxy in this lower right region. As for the probabilities of being spheroidal, in both bins they peak at the upper left corner, which is consistent with our intuition that galaxies are intrinsically rounder when we look at smaller objects. Such  probability maps can facilitate future morphological and kinematic observations aimed at searching for prolate galaxies at a range of redshifts, including at $z>3$ with \textit{James Webb Space Telescope} (\textit{JWST}). 

\begin{figure*}
\includegraphics[width=1.0\textwidth]{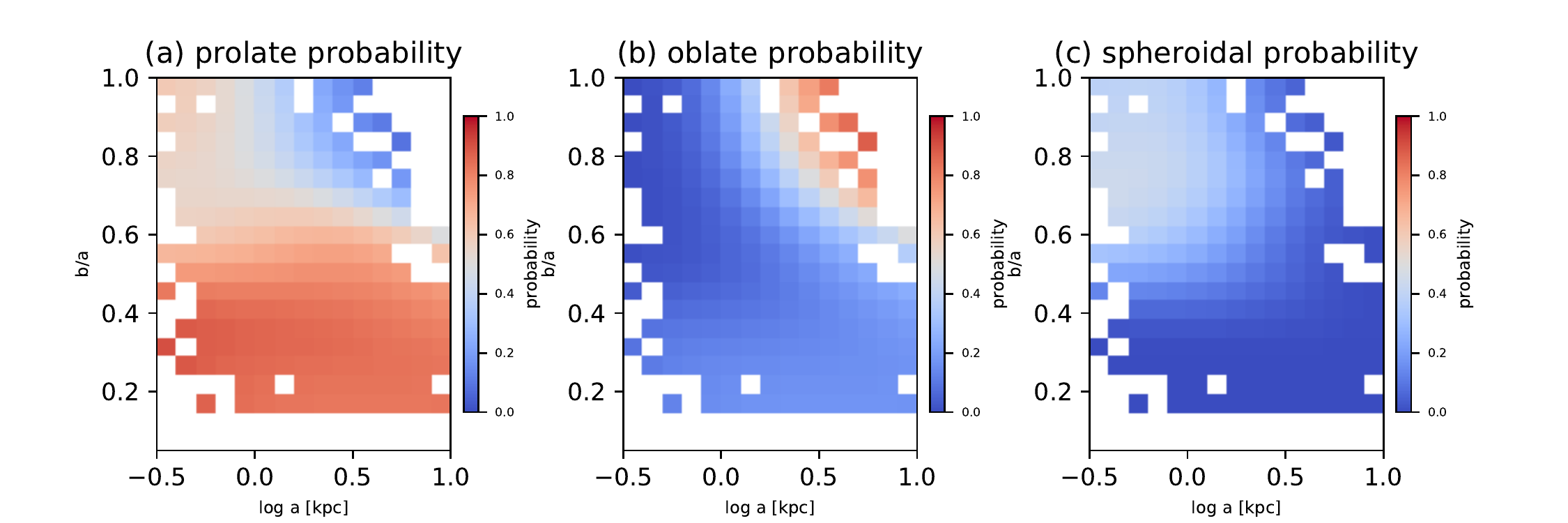}
\caption{The probability distribution of a CANDELS galaxy's being prolate, oblate or spheroidal over the $b/a-\log a$ plane for the early-prolate bin. Probabilities are only calculated in the bins containing at least one observed galaxy.}
\end{figure*}

\begin{figure*}
\includegraphics[width=1.0\textwidth]{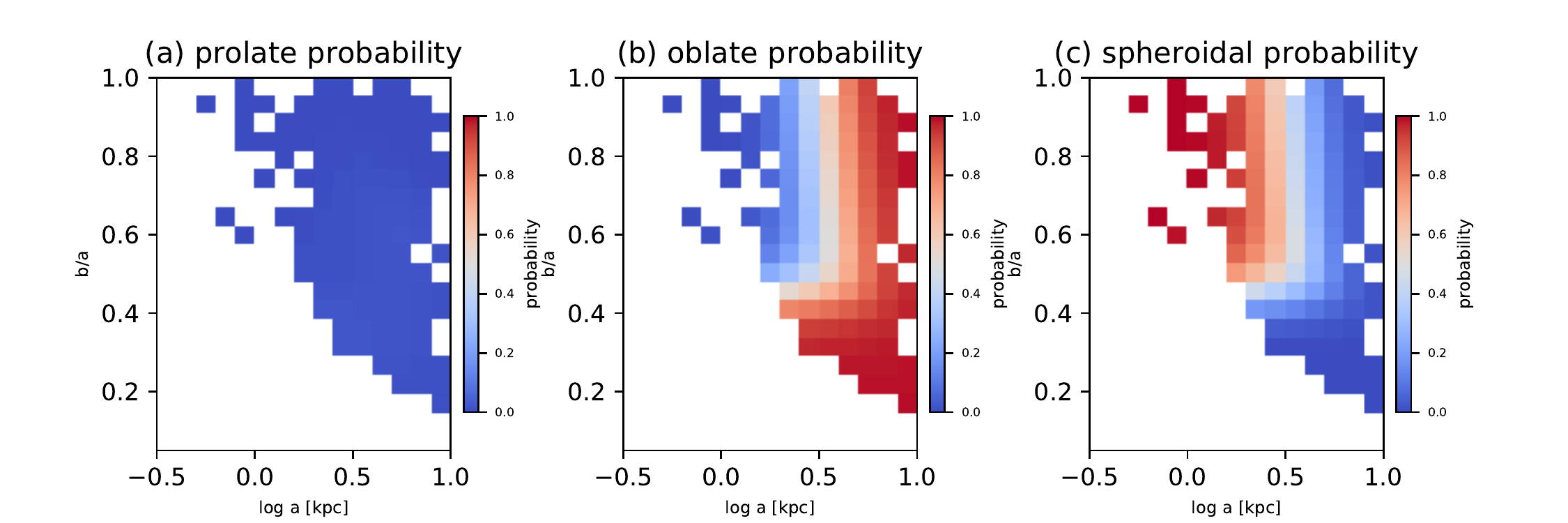}
\caption{The probability distribution of a CANDELS galaxy's being prolate, oblate or spheroidal over the $b/a-\log a$ plane, for redshift and mass interval for the late-oblate bin. Probabilities are only calculated in the bins containing at least one observed galaxy.}
\end{figure*}


\subsection{The modeling of the dust optical depth maps} 
\label{sub:the_modeling_of_the_dust_optical_depth_maps}

Another theoretical prediction we can make with such a modeling is the theoretical dust optical depth maps of galaxies on the $b/a-\log a$ plane. Such maps can be used as a sanity check of whether our modeling results are (qualitatively) consistent with the trends of $A_V$ with projected $b/a$ seen in \textbf{Fig. 1}.

\begin{figure}[htb]
\includegraphics[width=0.5\textwidth]{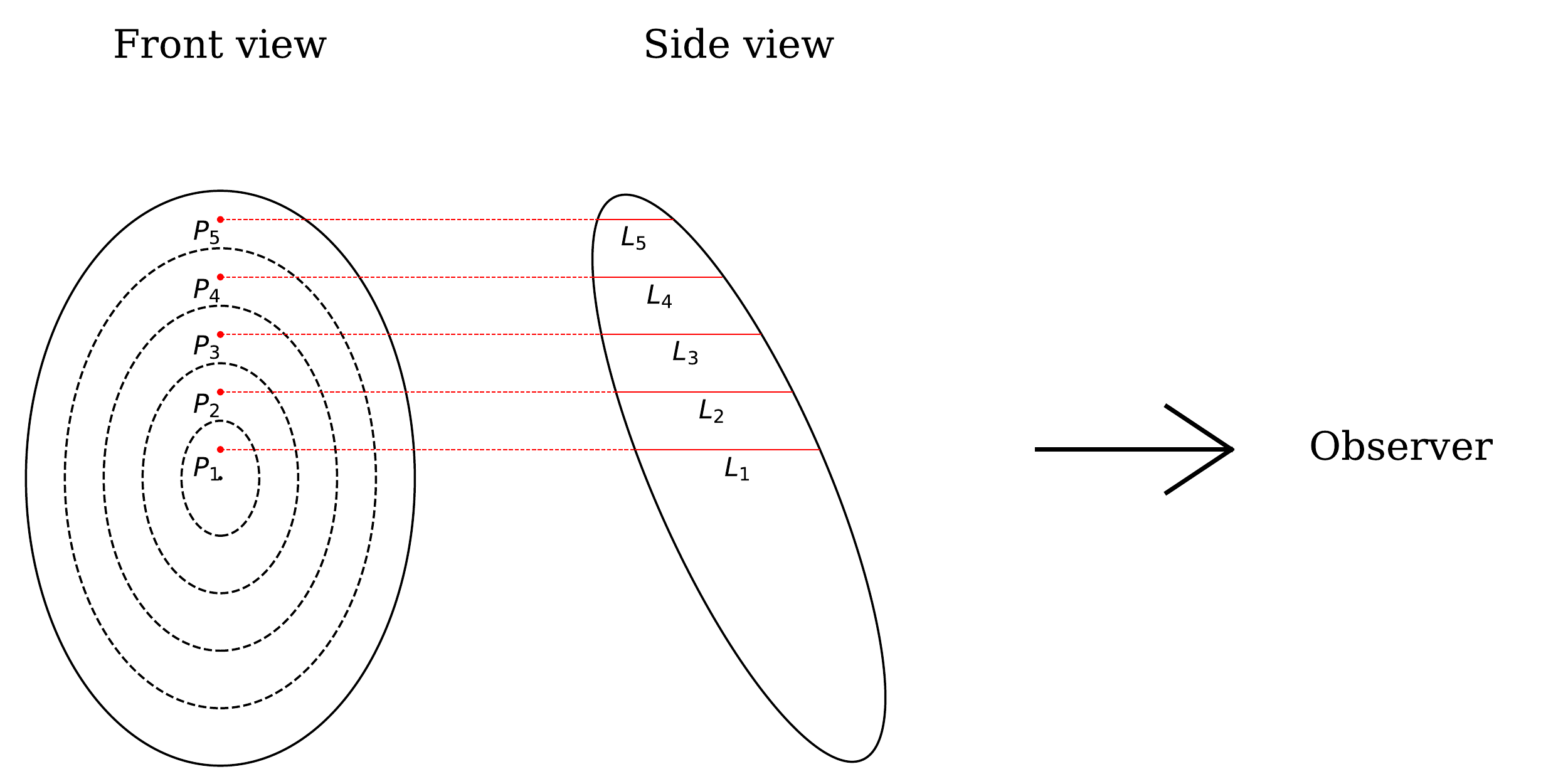}
\caption{Illustration of the calculation of the mean path lengths and $A_V$ values of model galaxies.}
\end{figure}

\begin{figure}[htb]
\centering

\subfigure[]{\label{fig:subfig:a}\includegraphics[width=0.5\textwidth]{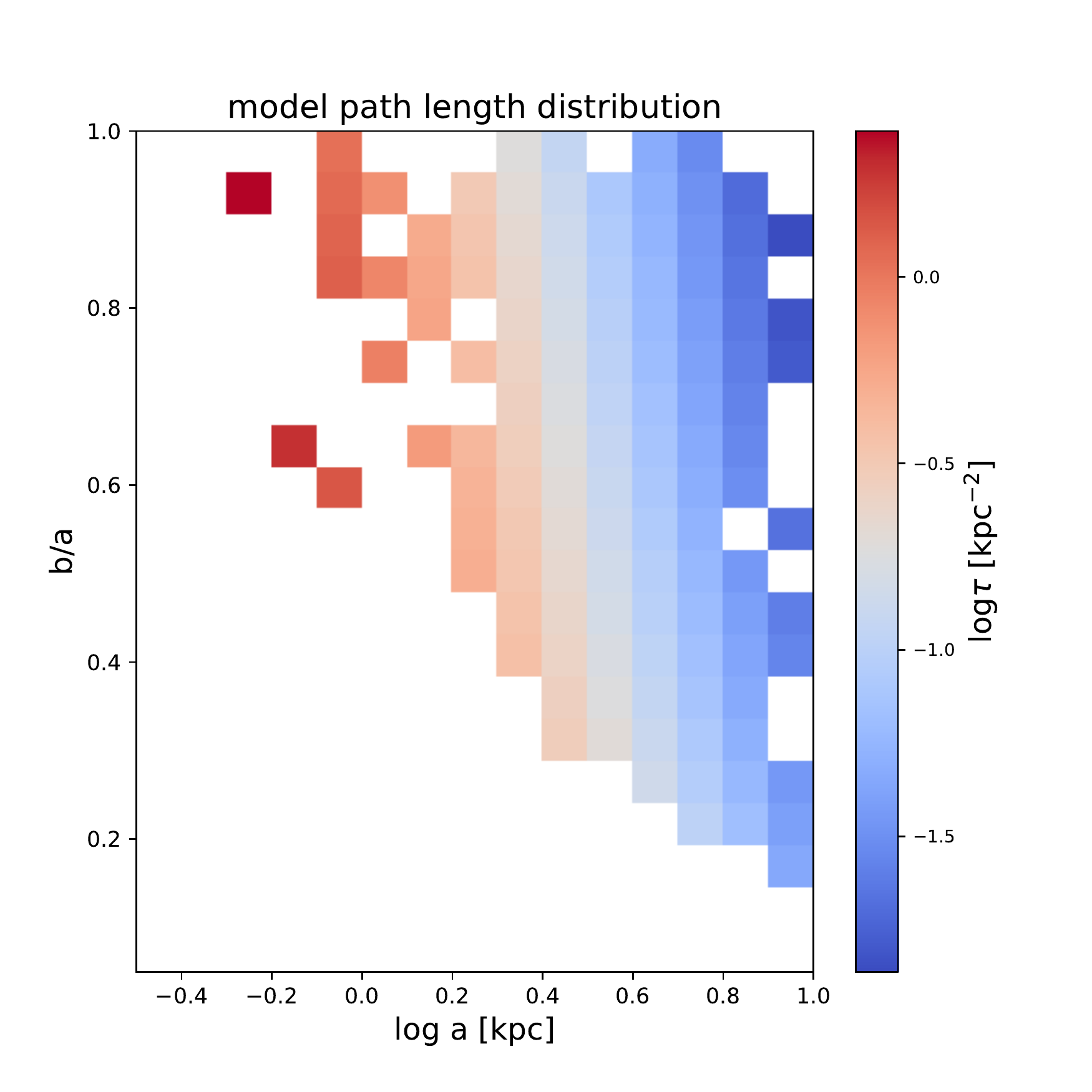}}
\hspace{0.01\linewidth}
\subfigure[]{\label{fig:subfig:b}\includegraphics[width=0.5\textwidth]{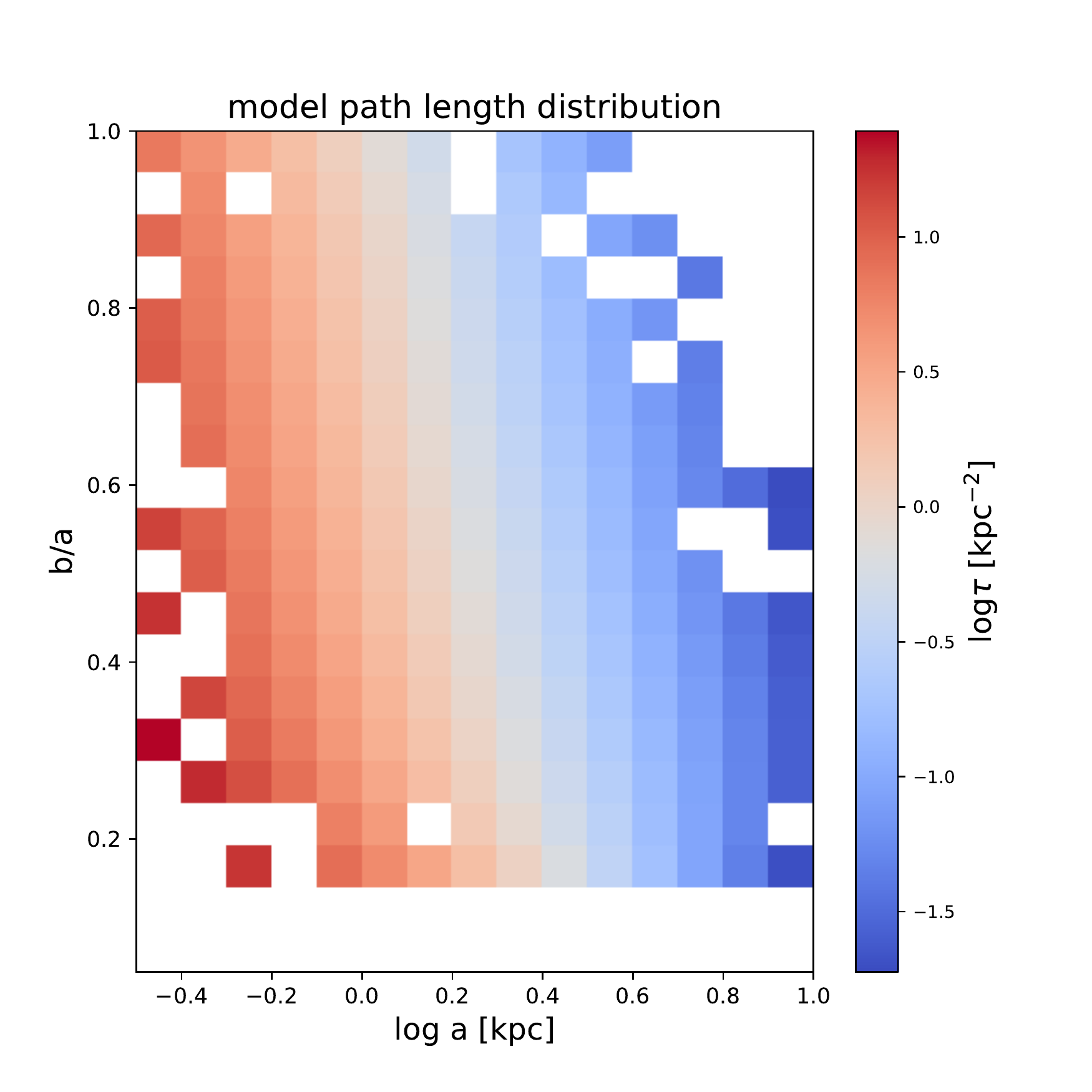}}

\caption{\textbf{Panel (a)}: The $\log \tau$ distribution over the $b/a-\log a$ plane for the late-oblate bin. \textbf{Panel (b)}: The same distribution but for the early-prolate bin. Note the vertical and horizontal decreasing trend of $\log \tau$ with increasing $b/a$ and $\log a$.} 
\end{figure}

\begin{figure}[htb]
\centering
\subfigure[]{\label{fig:subfig:a}\includegraphics[width=0.5\textwidth]{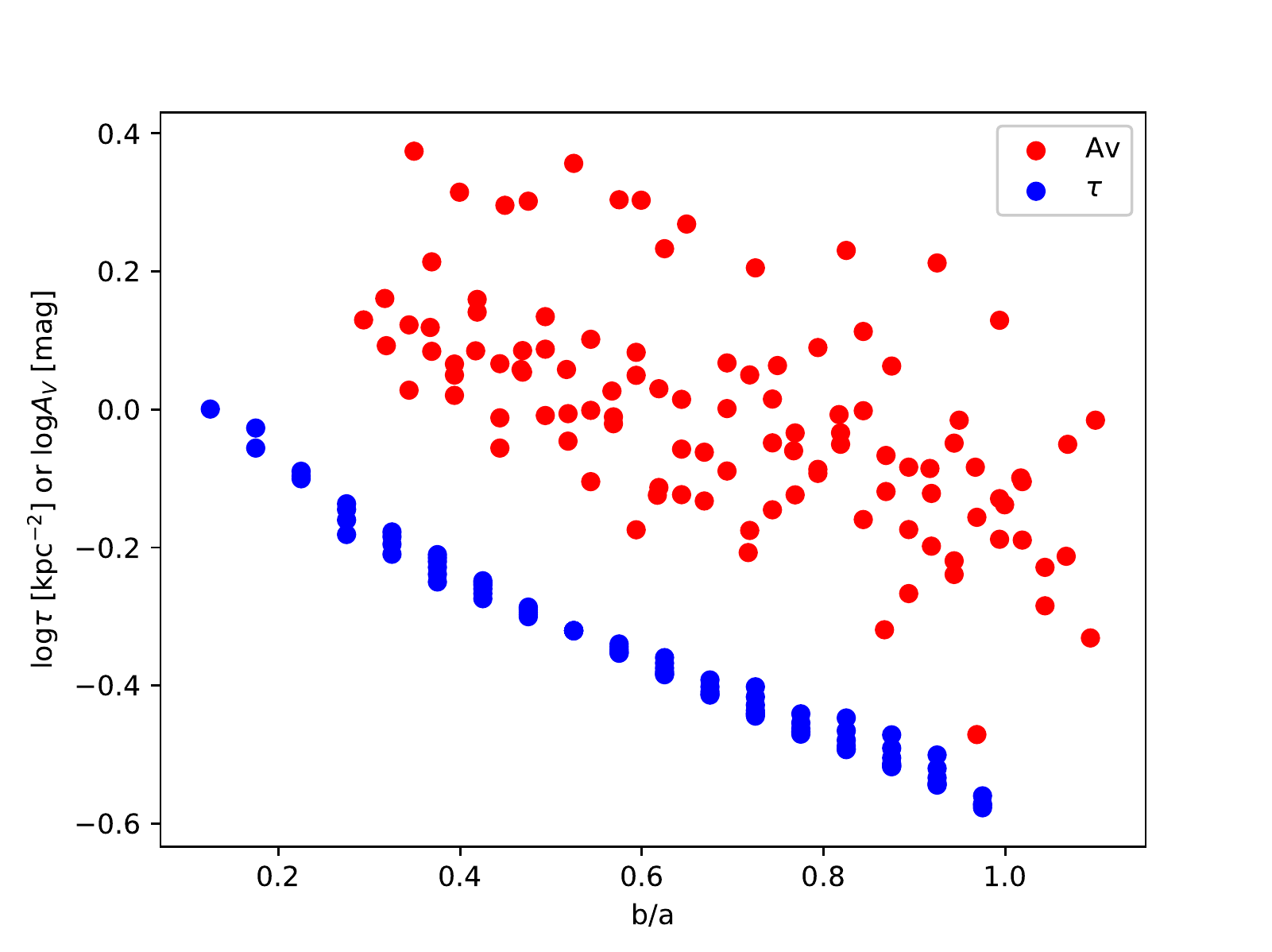}}
\hspace{0.01\linewidth}
\subfigure[]{\label{fig:subfig:b}\includegraphics[width=0.5\textwidth]{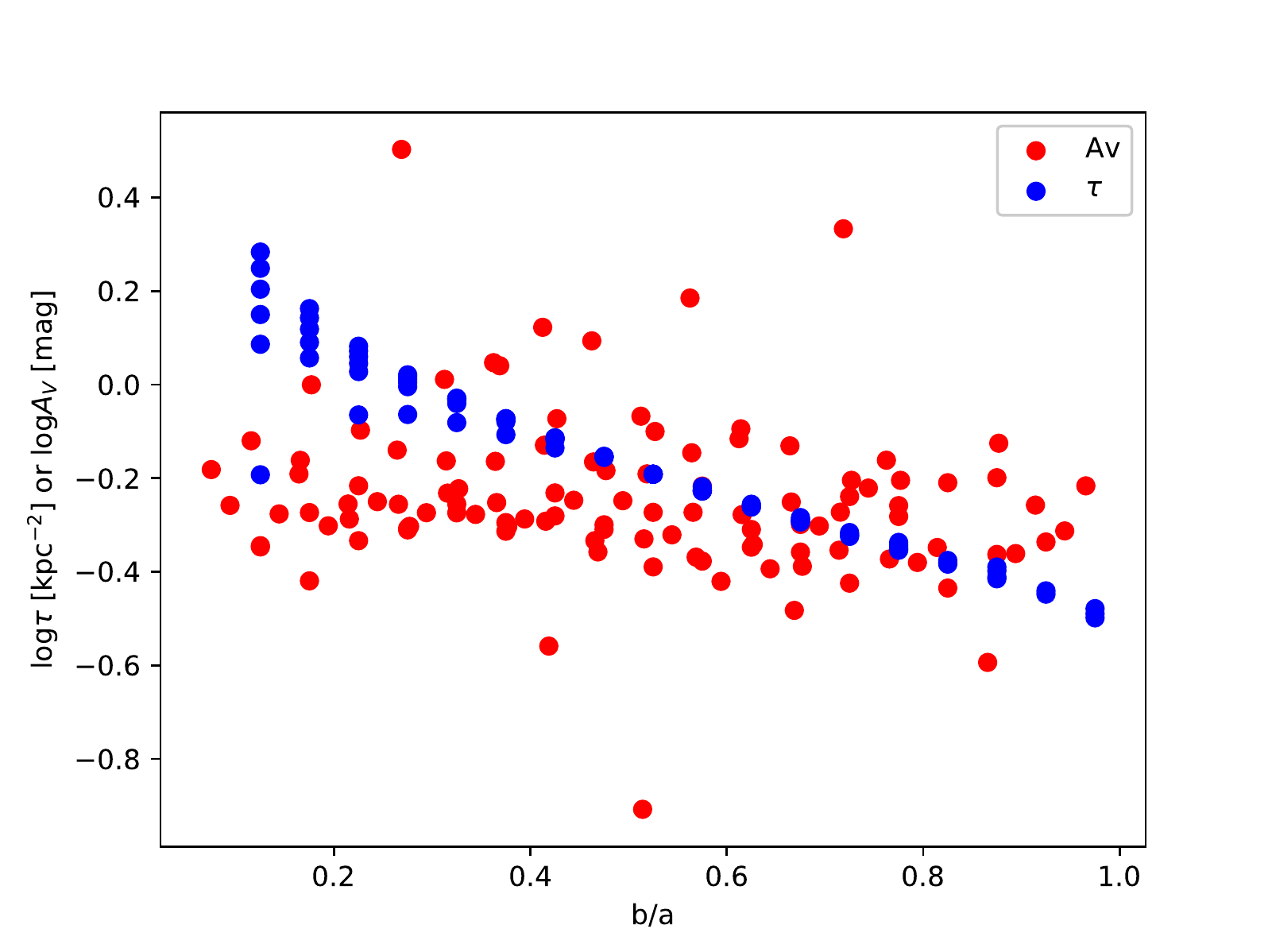}}
\caption{\textbf{Panel (a)}: The comparison between the aligned $\log A_V-b/a$ distribution and $\log \tau-b/a$ distribution of the galaxies in the late-oblate bin. \textbf{Panel (b)}: The same comparison but for objects in the early-prolate bin. Note that although the predicted $\log \tau$ curve are qualitatively consistent with the observed $A_V$ curve in the late-oblate bin, the opposite is the case for the early-prolate bin. The $\log \tau$ curve shows roughly the same slope in these two bins, which is a problem of our dust model.} 
\end{figure}

We first clarify what we are really modeling and support this choice with physical motivation. Ignoring scattering, the attenuation of starlight by interstellar dust, $\tau_{\nu}$, is:

\begin{equation}
\tau_{\nu} = \int n_{\mathrm{d}} \sigma_{\mathrm{ext}, \nu} dl\ ,
\end{equation}
where $n_{\mathrm{d}}$ is the volume number density of dust grains, $\sigma_{\mathrm{ext},\nu}$ is the extinction cross section at the frequency $\nu$, and $l$ is the path length. Therefore if we assume:

\begin{itemize}
\item[(1)] All the galaxies have the same number of dust grains;
\item[(2)] The composition and sizes of grains in all galaxies are identical;  
\item[(3)] The dust grains and stars are uniformly mixed within every galaxy.
\end{itemize}
then the optical depth at an arbitrary frequency is proportional to the mean path length $L$ through a galaxy divided by its total volume, i.e.:

\begin{equation}
\tau_{\nu} \propto \tau = \frac{L}{abc}\ .
\end{equation}

The $abc$ term in the denominator takes into account the fact that the dust density is smaller in larger-volume galaxies assuming total dust mass is constant.  In fact, CANDELS data show that dust mass is not constant -- smaller galaxies with lower projected a on average have less dust than larger galaxies (at fixed mass and redshift, Lin et al., in prep.). However, \textbf{Fig. 5} shows that galaxies of all shapes tend to appear in a narrow slice of projected $a$, i.e., that the amount of `$a$-crossing' due to projection effects is small, even for prolate and triaxial objects.  That being the case, it is appropriate to think of our optical depth maps as representing the variation of $A_V$ within a single slice of projected $a$, and this philosophy will be utilized in the analysis below.

Next we demonstrate our method to calculate the mean path length $L$. As shown in \textbf{Fig. 21}, we first divide the whole image\footnote{In the solid ellipsoid modeling of galaxies, such an image is simply the projected two-dimensional ellipse from an arbitrary direction.} with four concentric ellipses, the semi-major axes of which are $0.2,\ 0.4,\ 0.6,$ and $0.8R_e$, respectively. On each semi-major axis we pick 5 sample points, corresponding to $r=0.1,\ 0.3,\ 0.5,\ 0.7$ and $0.9R_e$, and calculate the path lengths, $L_i,\ i=1,2,...,5$, through these five points along the line of sight. We further assume that every point in each elliptical ring has the same path length as the corresponding sample point, and therefore the total mean path length of the projected galaxy is
\begin{equation}
L = \frac{\sum_{i=1}^5 L_i \cdot A_i}{\sum_{i=1}^5 A_i} \ ,
\end{equation}
where $A_i$ is the area of the i-th elliptical ring. As long as we get the path length for each galaxy observed from an arbitrary direction, we can immediately calculate the model optical depth map by calculating the mean optical depth in all the $b/a-\log a$ bins. 

\textbf{Fig. 22} shows the predicted mean optical depth map for the galaxies in both the late-oblate bin (panel (a)) and the early-prolate bin (panel (b)). There are some noteworthy features in these maps. Firstly, there's a clear trend that the mean optical depth increases as $b/a$ decreases in each slice of projected $a$, which is consistent with the increasing $A_V$ trend with decreasing $b/a$. Secondly, an even more pronounced trend is seen that as the size of the galaxy decreases (especially when the semi-major axis is smaller than 1kpc), the mean optical depth gets larger. Such a trend is seen in the actual data \citep[][Lin et al., in prep.]{FangThesis}, but it is smaller than seen in \textbf{Fig. 23}.  The large trend there stems from assuming that all galaxies in a given mass-redshift bin have the same amount of dust (Eq. 6), which evidently is not quite true -- smaller galaxies have somewhat less dust.

Nevertheless, as noted above, each type of galaxy (as given by its intrinsic $a,b,c$) tends to populate a single slice of projected $a$, and so we can tune our comparison in the following way: We align all the mean $\log A_V$-$b/a$ curves (in each $\log a$ slice) so that the mean positions $\left( \langle{b/a}\rangle, \langle{\log A_V}\rangle \right)$ coincide, and align the $\log \tau-b/a$ curves (in each $\log a$ slice) so that the $\tau$ values at $b/a\sim 0.5$ are the same. In this way, we effectively bypass the systematic difference between the total dust masses of galaxies with different $a$. \textbf{Fig. 23} shows such aligned curves (again for both the late-oblate bin and the early-prolate bin) \footnote{Figs. 22 and 23 illustrating all the modeled redshift and mass bins are available at \url{https://sites.google.com/site/zhw11387/Home/research}.}. We can see that for the late-oblate bin (\textbf{Fig. A1 (a)}), the agreement between the slope of the two curves (blue and red) is quite good. But the opposite happens in the early-prolate bin (\textbf{Fig. A1 (b)}), where $\log A_V$ remains roughly constant with the projected $b/a$, while $\log \tau$ shows similar value and slopes as in \textbf{Fig. A1 (a)}. This tells us that our dust model is unable to differentiate between the oblate-/prolate-dominated populations. The ultimate reason is that we have assumed that every galaxy has the same number of dust grains. Therefore, a pair of prolate galaxy and disk object with similar projected $(\log a, b/a)$ would have the same dust column density, which leads to the same $A_V$ value. To address the inconsistency between the predicted $\log \tau-b/a$ curve and the $\log A_V-b/a$ curve, we note that we have assumed that the gas and dust grains are uniformly mixed with the stellar components of a galaxy. But actually this may not be the case. \citet{Ceverino:2015db} has found that some of the VELA galaxies have a gas disk even when their stellar component shapes are prolate. Since the magnitude of dust attenuation scales with the amount of gas times metallicity, we will still observe a larger $A_V$ value when the {\it gas} disk is seen edge-on. Because of the weak alignments between the main axes of the gas and stellar components of galaxies in their prolate phase \citep{Tomassetti:2015ed}, it's not straightforward to predict the $A_V$ map by the projection of {\it stellar} mass profiles, therefore it's not surprising if we see no systematic $A_V$ trends with $b/a$ in redshift-mass bins dominated by prolate galaxies. On the contrary, after the galaxies become disky, the alignment between the gas disk and the stellar disk strengthens \citep{Tomassetti:2015ed}. In this case, our assumption becomes more realistic and therefore the result matches the data, which shows a strong correlation between $A_V$ and $b/a$. In short, the lack of correlation between $A_V$ and $b/a$ for prolate-dominated populations and strong correlation for oblate-dominated ones is consistent with predictions based on the shape evolution of different components of the VELA galaxies. They may therefore provide a useful framework for future dust analysis.

Another insight from the different shapes of the gas and stellar components of a prolate galaxy in VELA simulations is that the basic kinematic properties of gas in a prolate potential, unlike in an oblate one, is not understood from an observational perspective yet. To understand this aspect, dust should be used as an independent structural probe of the gas component of a galaxy. The Near-Infrared Spectrograph \citep[NIRSpec,][]{2012AAS...21924125F} of \textit{JWST} will be able to resolve both the emission lines that are tracers of gas and the absorption lines of stars, to study the kinematics of these two different components of galaxies separately. This will lead to a more thorough understanding of such differences in the kinematics of the gas and stellar components of prolate galaxies.


\section{Caveats} 
\label{sec:caveats}

In this section, we summarize a number of potential problems with the analysis in this paper.

Our modeling is based on the fundamental assumption that a galaxy can be modeled as a uniform three-dimensional ellipsoid, and when viewed from a certain direction, the observed $b/a$ and $a$ are calculated from the projection of that ellipsoid. A more realistic modeling is to model the galaxies as single S\'ersic light profiles and get their apparent $b/a-\log a$ distribution by generating randomly-viewed mock images with the method introduced by van de Ven \& van der Wel in prep., and measuring these images with \textsc{GALFIT}. See the Appendix for a discussion in more detail about such a S\'ersic modeling. Due to the limitation set by the available computational resources we are currently unable to calculate full $b/a-\log a$ distributions for these more realistic models. But based on the similarity between the distributions shown \textbf{in Fig. 5}, we argue that our main conclusions won't change qualitatively if we implement this more realistic S\'ersic modeling.

On the other hand, we have assumed that the galaxy population in each redshift and mass bin has a multivariate normal distribution in $\left(E, T, \gamma=\log a\right)$ parameter space. This assumption is a potential source of systematic effect that the modeled projected $b/a-c/a$ distributions don't look like that of the \emph{mass profiles} of the VELA galaxies, in the sense that the latter has many objects in highly spheroidal regions, and few galaxies in the intermediate region between prolate and settled galaxies as defined by \textbf{Fig. 17}, while the opposite is the case for the modeled distributions based on CANDELS data. In principle, it may also be the source of the redshift dependence of the transition mass range, at which a galaxy leaves the prolate phase and settles into disk galaxies, if the transition mass range is in fact independent of time, as is the case in VELA simulations. A potential way to address the difference between the modeled $b/a-c/a$ distributions and the one of the VELA galaxies (i.e. \textbf{Fig. 17}) is to add more multivariate normal populations in $\left(E, T, \gamma=\log a\right)$ parameter space. But given the limited sample size (which prohibits us from binning the data more finely) versus relatively large number of free parameters (eight more parameters for each additional multivariate normal population), we refrain from adding more of such populations. But our method is able and ought to be generalized into multiple population cases once there's adequate number of galaxies in each redshift-mass bin.

A third point is that our analysis method requires binning the data in two dimensions, which means that the numbers of objects in each bin are inevitably smaller than they are in one-dimensional $b/a$ modelings. In this work, a typical $b/a-\log a$ bin contains a few dozen objects, while in \citet{vanderWel:2014ka} it's common to see more than 100 galaxies in a single $b/a$ bin. This makes the results less statistically robust than the ones from previous work. Hopefully in the future it will be possible to get a larger sample of star-forming galaxies to address this problem. But given the fact that the modeled distribution captures the features of the observed data well, and that parameters vary smoothly and regularly between adjacent mass-redshift bins, we argue that our results are also quite robust.

Another caveat concerns the completeness of the galaxies in the highest redshift bins. According to our selection criterion based on total magnitudes, the samples are nominally complete in every mass-redshift bin.  However, in practice the catalogs may be incomplete near the magnitude limit due to loss of low-surface brightness galaxies. We believe this fact explains why, in Table 1, for the two less massive bins, the number of galaxies in the highest redshift bin is smaller than that in the next highest redshift bin. But we argue that this fact doesn't qualitatively change the main conclusions of this work. Firstly, even if we remove all the highest redshift bins, the trend that prolateness decreases with time and mass remains invariant. Secondly, if at high redshift, the populations of galaxies were dominated by oblate objects, we would be more likely to observe more small and edge-on disks rather than small and face-on ones, as the former have larger surface brightness. In fact what we observe is the existence of prolate galaxies in the high redshift bins. This would lead to the conclusion that the real curved boundaries are even more pronounced than what we observed in the highest redshift bins of \textbf{Fig. 1}. Therefore, no matter whether we take the incompleteness in the highest redshift bins into consideration, the qualitative conclusion that these bins are dominated by prolate objects still holds, although quantitatively the actual fractions of prolate galaxies may change.

A third caveat involves dust attenuation. As \citet{Padilla:2008gw} pointed out, the existence of dust may affect the detectability of disk galaxies in certain orientations. In the calculation of the apparent $b/a-\log a$ distribution of a galaxy, we assume that the galaxy is viewed from every direction in 4$\pi$ solid space with equal probability. But, due to the fact that the light from stars in a disk galaxy is absorbed more significantly when it is viewed edge-on, some galaxies may drop out of the detection limit when viewed in this direction. Therefore the true distribution will deviate from the one calculated in this work. 

The final point is that the dust model in this paper is not fully realistic. As is discussed in Section 5.3, our model may be unrealistic in the sense that it doesn't take into account the effect when the optical depths are high \citep[e.g.,][]{SeonDraine2016}, or the fact that the gas component may have a different shape from the stellar component and they are not mutually aligned \citep{Ceverino:2015db,Tomassetti:2015ed}.

\section{Conclusion} 
\label{sec:conclusion}

We have found that the shapes of the star-forming galaxies in CANDELS are correlated with their sizes, in the sense that smaller galaxies are intrinsically rounder. Motivated by this insight, we expanded the previous work by \citet{vanderWel:2014ka} by analyzing the projected $b/a-\log a$ distributions of the CANDELS star-forming galaxies with $0 < z < 2.5$ and $9 < \log \left(M_{*} / M_\odot\right) < 10.5$ in a grid of redshift and mass bins, assuming that the shapes of the three-dimensional light distributions of galaxies can be approximated by uniform triaxial ellipsoids, and the size and shape parameters of galaxies $\left(E, T, \gamma=\log a\right)$ has a multivariate normal distribution. By doing this modeling we give the fractions of prolate, oblate and spheroidal galaxies in each redshift and mass bin. Based on these fractions we find that galaxies tend to be prolate at low mass and high redshift, and oblate at high mass and low redshift, which means galaxies evolve from prolate to oblate. This transition tends to occur in a characteristic mass range, which tends to decline in time: roughly from $\log \left(M_{*} / M_\odot\right)\sim10.3$ at $z\sim2.25$ to $\log \left(M_{*} / M_\odot\right)\sim9.3$ at $z\sim0.75$. Qualitatively our findings, summarized in \textbf{Fig. 14}, are in line with those of \citet{vanderWel:2014ka}. But quantitatively we find more prolate and/or spheroidal galaxies than they did, due to the existence of a new correlation between the (intrinsic or projected) shapes and the sizes of galaxies, in the sense that smaller galaxies are systematically rounder. We also verify that such a correlation is not an artificial effect induced by selection bias in the CANDELS pipelines. The fact that we find more oblate objects in more massive and lower redshift bins is also consistent with the findings by \citet{Jiang2018}, who also effectively considered the correlation between the galaxies' intrinsic shapes and their sizes. But we note that when comparing small and large star-forming galaxies in the same redshift-mass bins, they argued that the latter are more likely to have disk-like components flattened by rotation. This is not fully consistent with our finding, which is that larger galaxies could be more disky \emph{or more prolate}. This is due to the fact that \citet{Jiang2018} didn't take into account the fact that prolate galaxies could also have large ellipticities, and thus not being able to cover the case in which some of the larger star-forming galaxies are in fact prolate, instead of disky.

We compared the results of the modeling of CANDELS data with the VELA simulation \citep{2014MNRAS.442.1545C,Zolotov:2015hk} data as `observed' in two-dimensional projection through dust, mimicing the observational features of the CANDELS observations. By comparing multi-band images from both datasets, we argue that it's not feasible to tell whether a galaxy is an edge-on disk or a prolate object from direct images due to their similar projected morphologies. Secondly, we demonstrate that in high-redshift and low-mass bins, the $b/a-\log a$ distributions are qualitatively consistent with a prolate-dominated galaxy population, by comparing CANDELS data in such bins with the same distributions of the prolate galaxies in the VELA simulation. Thirdly, we investigate the time evolution of the fractions of different shapes of galaxies in the VELA simulations and find the same trend with time and mass as found in the CANDELS data. This finding further confirms the picture proposed by \citet{Kassin:2012kz} and \citet{Simons:2016bk,Simons:2017bc}, based on kinematic data of real observed galaxies, and \citet{Kassin:2014jp,Ceverino:2015db,Tomassetti:2015ed,Ceverino:2017im}, based on hydrodynamic models, that the rotational support in galaxies grows with time, which is a process that starts earlier in more massive galaxies. Our results are also consistent with the predictions from the VELA simulations \citep{Ceverino:2015db,Tomassetti:2015ed} that the transition of shape occurs in a characteristic mass range, where galaxies tend to undergo a process of wet compaction to a blue nugget, and make a transition from being dark-matter dominated to baryon dominated. See also Huertas-Company et al. in prep. and Dekel et al. in prep..

How galaxies achieve their final structure is one of the most basic aspects of galaxy evolution.  The emerging story -- coming from both data and theory -- seems to be that the process is gradual but is more advanced in massive galaxies at a given redshift.  In this sense, the process of structural evolution is coming to resemble the cycle of star-formation and quenching, which is also more advanced in massive galaxies at each $z$.  This latter phenomenon has come to be known as `downsizing' \citep{Cowie:1996fb}. Drawing the parallel with star formation, \citet{Kassin:2012kz} coined the term `kinematic downsizing' to describe their finding that massive galaxies at any given redshift are more settled.  It will be interesting and instructive to compare and contrast the phenomena of structural evolution and the star-formation life-cycle going forward to see if their trajectories are in fact parallel and how their underlying physics compares.

Using the results of such modelings, we are able to give the probabilities of a galaxy's being prolate, oblate, or spheroidal, as a function of its redshift, mass, $b/a$ and $\log a$. This can be used to facilitate the target selections for kinematics or spectroscopic observations of the prolate galaxies in the future. We can also calculate the mean optical depth map across the $b/a-\log a$ plane. We find that in each $\log a$ slice, the trend of the mean optical depth and the $A_V$ value with $b/a$ are qualitatively consistent with each other in the late-oblate bin. The opposite happens for the early-prolate bin, possibly because the gas and stars are not aligned in prolate galaxies, with the gas mostly supported by rotation but the stars are supported by anisotropic velocity dispersion.

For readers' convenience we also summarize the availability of the data and results involved in this paper here. The CANDELS catalog can be accessed via \url{http://arcoiris.ucsc.edu//Rainbow\_navigator\_public/}. The ETa modeling results given by this work, including the comparisons between the data and model, the probability (see Section 5.2) maps and the model optical depth maps (see Section 5.3), can be found at \url{https://sites.google.com/site/zhw11387/Home/research}.


\section*{Acknowledgements}

This work is partly funded by China Scholarship Council. HZ is grateful for financial support from Mr. Mingzhi Li and the Fujian Fuguang Foundation, and helpful discussions with Dr. Bradford Holden and Mr. Viraj Pandya from UCO/Lick Observatory, Dept. of Astronomy and Astrophysics, University of California, Santa Cruz, CA 95064 USA, and Mr. Haobin Lin from Peking University, PRC. The VELA simulations were run on NASA's Pleiades supercomputer at NASA Ames Research Center and on NERSC supercomputers at Lawrence Berkeley National Laboratory.  JRP acknowledges support from grant HST-AR-14578.001-A. AD, SMF and JRP acknowledge support from the US-Israel BSF 2014-273. AD and AvW acknowledge support from GIF I-1341-303.7/2016. AD is partly supported by NSF AST-1405962 and DIP STE1869/2-1 GE625/17-1. Support for Program number HST-AR-15025 was provided by NASA through a grant from the Space Telescope Science Institute, which is operated by the Association of Universities for Research in Astronomy, Incorporated, under NASA contract NAS5-26555. Support for the CANDELS survey, HST-GO-12060, was provided by NASA through a grant from the Space Telescope Science Institute, which is operated by the Association of Universities for Research in Astronomy, Incorporated, under NASA contract NAS5-26555. SMF and DCK acknowledge support from NSF-AST-1615730. DC has been funded by the ERC Advanced Grant, STARLIGHT: Formation of the First Stars (project number 339177). The VELA simulations were performed at the National Energy Research Scientific Computing Center (NERSC) at Lawrence Berkeley National Laboratory, and at NASA Advanced Supercomputing (NAS) at NASA Ames Research Center.




\bibliographystyle{mnras}
\bibliography{Prolate} 




\appendix
\section{Tests of Potential Selection Effects in the CANDELS Pipeline} 
\label{sec:tests_on_the_potential_selection_effects_in_candels_pipeline}

In Section 2 we claimed that the curved boundary and the lack of objects in the upper right corner of each $b/a-\log a$ panel is not due to the selection effects induced by the detection scheme in CANDELS or the measurement process in \textsc{GALFIT}. Here we describe how a two-step experiment using SExtractor and \textsc{GALFIT} \citep{Peng:2010eh} establishes this by mocking up the procedures implemented in the pipeline.

We first explain the possible reasons why these two features could be induced by the selection bias in source detection and measurements. As pointed out in Section 3, the probability distribution of a prolate galaxy in $b/a-\log a$ space is a curved ridgeline from lower right to upper left in this space and is thus naturally compatible with these two features; on the other hand, the distribution of a spheroidal galaxy is localized at a high $b/a$ value, and thus the detectability of such an object and the accuracy of the measurements wouldn't change much as the viewing direction varies. Therefore, it's natural to conclude that the selection bias of the pipeline, if any, would affect the oblate populations in the sample most significantly.

Bearing this in mind, it's easy to imagine how the two features in the observed distribution can be explained by a selection bias on oblate objects:

\begin{itemize}
\item[(1)] The lack of objects in the upper right corner may be due to the fact that, when a disk galaxy is viewed face-on, due to the smaller path length through the galaxy, its surface brightness is lower than when it is viewed edge-on. This decrease of surface brightness may make some of the fainter disk galaxies drop out of the sample because they're too faint to be detected when viewed face-on. But their edge-on counterparts are still included in the sample due to relatively higher surface brightness. Thus this kind of discrepancy of different surface brightness at different $b/a$ values could cause the lack of objects in the upper right corner.
\item[(2)] The curved lower boundary essentially tells us that there is minimum shortest main axis length of galaxies which is not proportional to the semi-major axis, and therefore we don't observe many small and highly flattened objects. But in principle this may simply be an artificial effect induced by either the detection or the measurement scheme. Firstly, when a small galaxy gets more and more flattened (and correspondingly fainter and fainter, which can be seen from the visual inspection of the images), its image gets more and more vulnerable to being lost in sky noise. Given the dual-mode SExtractor detection strategy \citep{Galametz:2013dd}, it is possible that such an object escapes detection in the cold mode and ends up being separated into multiple objects in the hot mode. In this case this galaxy would be regarded as multiple objects, so naturally we can't get small and also very flattened galaxies in the catalog. Another possible cause lies in the measurement phase. As we know all the images are pixelated, and thus the fact that we didn't observe such small and flattened objects may be due to the fact that these objects have a too small semi-minor axis which is significantly smaller than 1 pixel, so that all the measurements on such objects are not usable.
\end{itemize}

Based on these hypotheses for the origin of these two features, we do the following experiments:

\begin{itemize}
\item[(1)] To investigate the origin of the lack of objects in the upper right corner, we pick out large and thin edge-on disk galaxies from the catalog, calculate their corresponding face-on light profiles using the formalisms introduced by van de Ven \& van der Wel in prep., add realistic sky background noise, feed these mock images into SExtractor and \textsc{GALFIT}, and see whether we get fewer objects detected or measured accurately. The setups of SExtractor and \textsc{GALFIT} are identical to those introduced in \citet{Galametz:2013dd} and \citet{2012ApJS..203...24V}.

\item[(2)] To investigate the origin of the curved lower boundary, we pick out small and round galaxies and calculate their edge-on light profiles based on the assumption that they are small and thin disks. After that all the procedures are the same as with experiment (1).
\end{itemize}

\begin{figure*}
\centering
\includegraphics[width=1.0\textwidth]{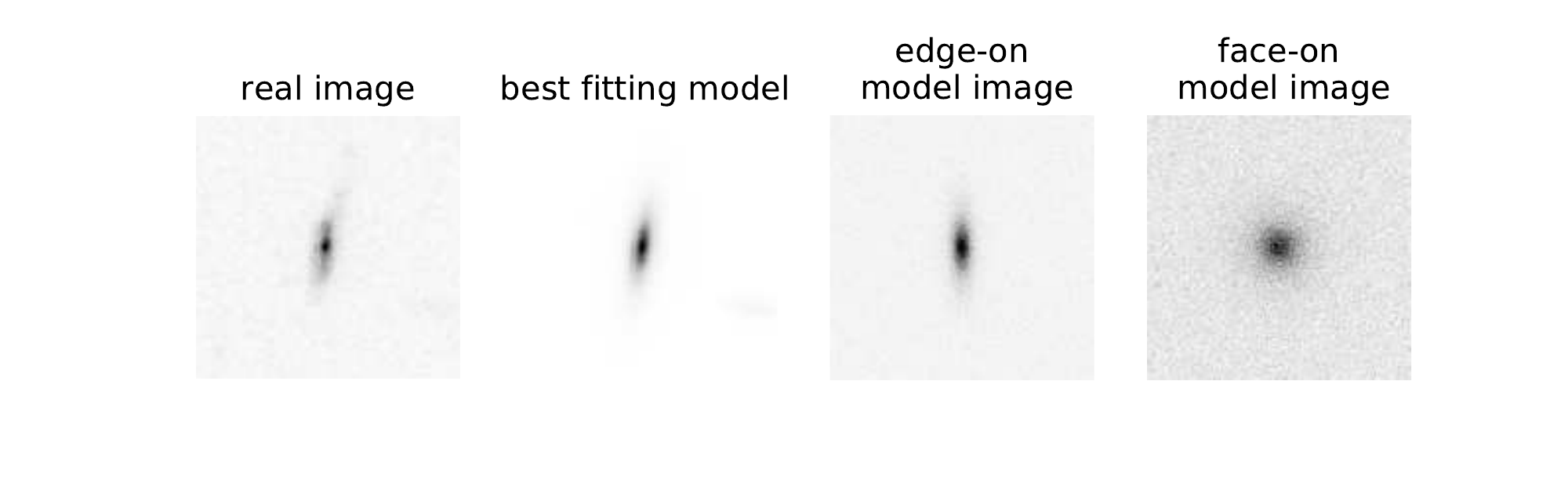}
\caption{Illustration of mock image generation. The first panel from the left is the cutout stamp from a CANDELS multidrizzled image. The second panel is the best fitting S\'ersic model obtained by \citet{2012ApJS..203...24V}. The third panel is the edge-on mock image generated using the formalisms from van de Ven \& van der Wel in prep., based on the assumption that the galaxy is oblate, and with PSF and realistic sky background noise taken into account. The fourth panel is generated in the identical way to the third one, except for being viewed face-on.}
\label{figA1}
\end{figure*}
\vskip 0.1in

\textbf{Fig. A1} shows an example of such a deprojected S\'ersic modeling. From left to right, the figure shows the original image from CANDELS, the best fitting model from \textsc{GALFIT}, the edge-on model image, and the face-on model image. By doing such experiments, we found that large and round (small and thin) disks do have lower detectability than their large and thin (small and round) counterparts, but only roughly by $\sim 20\%$, which is far from sufficient to account for the lack of objects at either the upper right or the lower left corner. On the other hand, when we view the small model disk galaxies edge-on, their $b/a$ can still be measured accurately. Therefore we arrive at the conclusion that neither of these two features is caused by the selection bias intrinsic to the CANDELS pipeline.

\section{Single-S\'ersic Modeling of Galaxies} 
\label{sec:single_sersic_modeling_of_galaxies}

Throughout this work we have been modeling the galaxies as solid triaxial ellipsoids (ellipsoidal modeling), with projected structural parameters measured directly from their two-dimensional projections. In fact there is a potential improvement of such a kind of modeling, which will be introduced in this section.

According to van de Ven \& van der Wel in prep., every two-dimensional S\'ersic profile can be produced by projecting a three-dimensional profile (hereafter 3D profile) in a certain direction. Thus it's feasible for us to model every galaxy as such a 3D profile. For different galaxies, the scale lengths and the shapes of the 3D profiles differ. By the projection of each of these galaxies randomly in many directions we get multiple images, which will be fed into \textsc{GALFIT} to measure their structural parameters. As a result, we get a $b/a-\log a$ distribution measured by \textsc{GALFIT}, instead of a simple geometrical calculation, and we can use these distributions in the modeling instead.

In principle this S\'ersic modeling method is more realistic than the solid ellipsoid method, in the sense that firstly it assumes a galaxy can be well approximated by a single two-dimensional S\'ersic profile, which, although still simplified, is much closer to the reality than the ellipsoidal modeling, in the sense that the galaxies are transparent in this model. Secondly the $b/a$ and $a$ come from \textsc{GALFIT}, which is the exact way of getting these parameters in real observations. However, such a method has the unfortunate drawback that to generate as many element distributions as we did with the solid modeling, the measurements of \textsc{GALFIT} would take more computer time than is available for us at the moment. Another possibility is to make use of a deep learning technique, e.g. \citet{Tuccillo:2017tq}, which is demonstrated to be much faster than \textsc{GALFIT}, to carry out massive measurements like this. But it remains to be evaluated whether such deep learning measurements are statistically consistent with what one gets from \textsc{GALFIT}.

As is seen in \textbf{Fig. 5}, the $b/a-\log a$ distributions generated by the two modeling methods are qualitatively similar in their shapes, but the magnitudes of scatter differ. This may be due to the fact that the measurement uncertainties obtained by \citet{2012ApJS..203...24V} incorporate random measurement uncertainties by comparing the measurements for the same objects in different data sets, which is not implemented in our S\'ersic modeling. But the overall similarities between the shapes of the distributions from the two different modelings convinces us that the main conclusions in this work won't change qualitatively if we adopt the more realistic S\'ersic modeling instead.

\section{The Effect of Empirical Corrections to the $b/a-\log a$ Distributions} 
\label{sec:the_effect_of_empirical_corrections_to_the_}

As was pointed out in Section 4.1, the effect of dust on the measurements of the semi-major axis $a$ of edge-on oblate galaxies ought to change the shape of the distributions, producing small tails in the lower right corner. We tried to make some empirical corrections to this in some of the redshift-mass bins. From \textbf{Fig. 1} we see that the only two bins where such tails show up significantly are $0.5 < z < 1.0$ with $9.5 < \log \left( M_{*} / M_\odot\right) < 10.0$ and $10.0 < \log \left( M_{*} / M_\odot\right) < 10.5$. In these two bins we assume the lower right tails are dominated by oblate galaxies that have been moved to larger a and smaller b/a due to central dust attenuation. According to \textbf{Fig. 5}, we know that the $b/a-\log a$ distribution of a disk galaxy should possess a vertical boundary; thus we move the data points leftward in the tails so that the corrected boundary is roughly a vertical line. In the correction we assume the semi-minor axis $b$ of the galaxy is invariant; therefore a correction of $a$ naturally leads to a correction of $b/a$. As a result, the data points are moved diagonally upward instead of horizontally. \textbf{Fig. C1} compares the corrected and uncorrected distributions of the galaxies with $0.5 < z < 1.0$ and $10 < \log \left( M_{*} / M_\odot\right) < 10.5$ (i.e. the late-oblate bin). 

\begin{figure*}
\centering
\subfigure[uncorrected]{\label{fig:subfig:a}\includegraphics[width=0.45\textwidth]{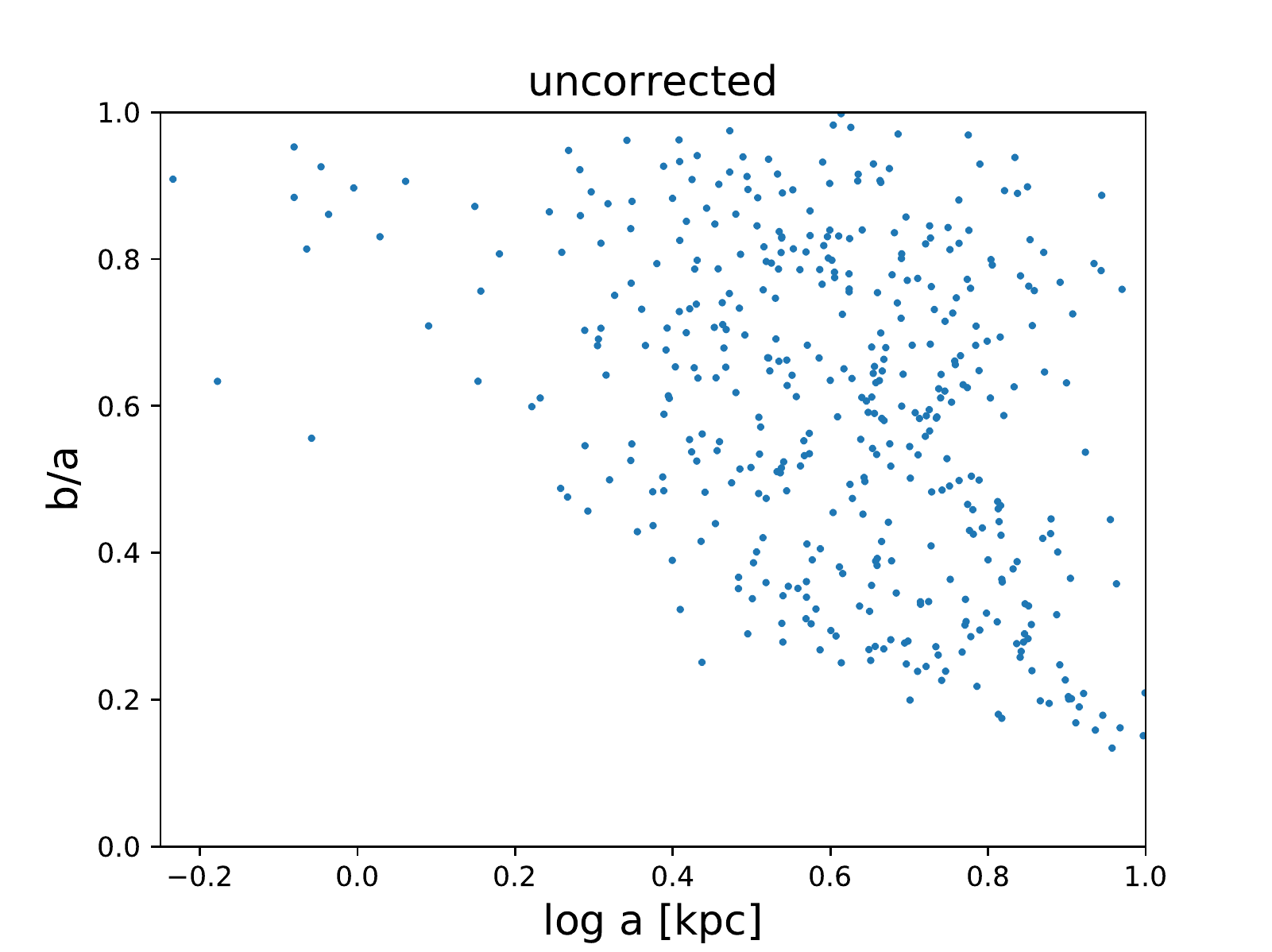}}
\hspace{0.01\linewidth}
\subfigure[corrected]{\label{fig:subfig:b}\includegraphics[width=0.45\textwidth]{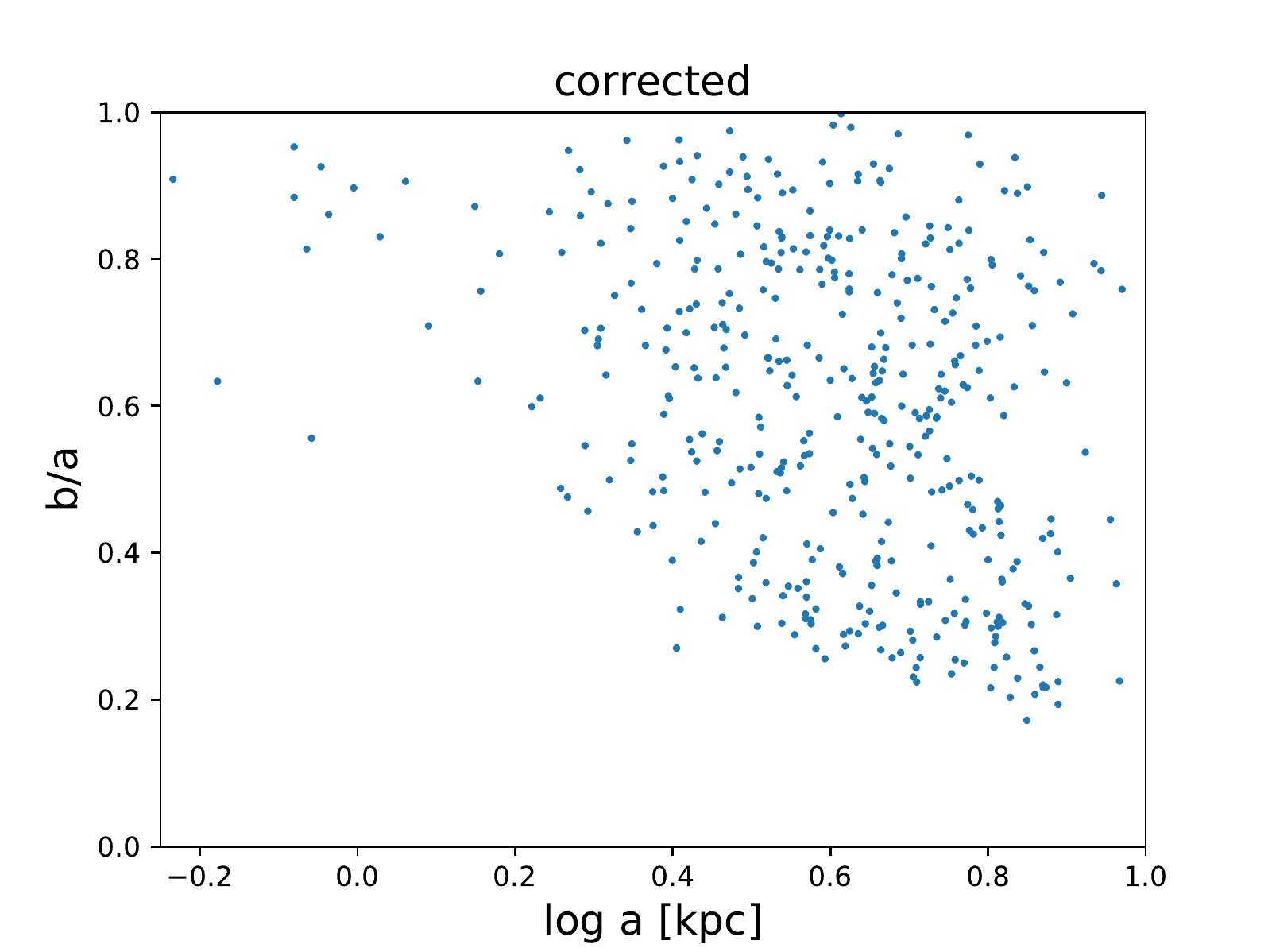}}
\centering
\caption{The uncorrected and corrected $b/a-\log a$ distribution of the galaxies with $0.5 < z < 1.0$ and $10 < \log \left( M_{*} / M_\odot\right) < 10.5$ (i.e. the late-oblate bin). Left panel: the uncorrected distribution. Right panel: the corrected distribution.} 
\end{figure*}

\begin{table*}
\begin{tabular}{cccccccccccccc}
\hline
\hline
Redshift & $\mathrm{log}\left(M_{*}/M_\odot\right)$ & correction &$\bar{E}$ & $\bar{T}$ & $\bar{\gamma}$ &
$\sigma_{EE}$ & $\sigma_{TT}$ & $\sigma_{\gamma\gamma}$ &
$\sigma_{E\gamma}$ & $f_{\mathrm{prolate}}$ &
$f_{\mathrm{oblate}}$ & $f_{\mathrm{spheroidal}}$ & $N_{\mathrm{obs}}$\\
\hline
$0.75$ & $9.75 $ & no & $0.745$ & $0.156$ & $0.574$     &   $0.008$ &  $0.658$ & $0.043$  & $0.016$ & $0.324$ & $0.561$ & $0.115$ & 1024  \\
$0.75$ & $9.75 $ & yes & $0.747$ & $0.174$ & $0.567$     &   $0.004$ &  $0.627$ & $0.039$  & $0.010$ & $0.347$ & $0.599$ & $0.053$ & 1024   \\
$0.75$ & $10.25$ & no & $0.728$ & $0.166$ & $0.680$    &   $0.035$ &  $0.039$ & $0.062$  & $0.038$ & $0.013$ & $0.678$ & $0.309$ & 426   \\
$0.75$ & $10.25$ & yes & $0.714$ & $0.133$ & $0.652$    &   $0.011$ &  $0.039$ & $0.054$  & $0.022$ & $0.012$ & $0.802$ & $0.186$ & 426
\end{tabular}
\caption{Best fitting model parameters and fractions of the three shapes of each redshift-mass bin.}
\end{table*}

We fed the corrected distribution to our modeling code to see whether the results changed significantly. The comparison between the best fitting parameters and the fractions of the different shapes is listed in \textbf{Table C1}. From the comparison we can see that we find $6.8\%$ ($18.3\%$) more oblate galaxies in the corrected distribution in the $0.5<z<1$ and $9.5 < \log \left( M_{*} / M_\odot\right) < 10$ ($10 < \log \left( M_{*} / M_\odot\right) < 10.5$) bin, which is expected because the corrected $b/a-\log a$ has a roughly vertical boundary, characteristic of an oblate population. But qualitatively our picture that the oblateness increases with time and mass is not only not damaged by this correction but instead strengthened by the larger fractions of oblate galaxies found in the low redshift universe. Therefore we argue that such empirical corrections don't affect the conclusions of this work.

\vskip 0.2in






\bsp	
\label{lastpage}
\end{document}